\documentclass[aps,showpacs,twocolumn]{revtex4}
\usepackage{graphicx}

\begin{document}

\newcommand{\lw}[1]{\smash{\lower2.ex\hbox{#1}}}


\title{
Spacetime structure of static solutions
in Gauss-Bonnet gravity: neutral case
} 

\author{
Takashi Torii$^{(a)}$, and Hideki Maeda$^{(b)}$
}

\address{ 
$^{(a)}$Graduate School of Science,
Waseda University, Shinjuku-ku, Tokyo 169-8555, Japan
}
\address{
$^{(b)}$Advanced Research Institute for Science and Engineering,
Waseda University, Shinjuku-ku, Tokyo 169-8555, Japan
}

\date{\today}

\begin{abstract}
We study the  spacetime structures of the static solutions in
the $n$-dimensional Einstein-Gauss-Bonnet-$\Lambda$ system systematically. 
We assume the Gauss-Bonnet coefficient $\alpha$ is non-negative and a cosmological constant is either positive, zero, or negative.
The solutions have the $(n-2)$-dimensional Euclidean
sub-manifold, which is the Einstein manifold with the curvature $k=1,~0$ and $-1$.
We also assume $4{\tilde \alpha}/\ell^2\leq 1$, where $\ell$ is the curvature radius, in order for the sourceless solution  ($M=0$) to be defined.
The general solutions are classified into plus and minus branches. 
The structures of the center, horizons, infinity and the singular point depend on the parameters $\alpha$, $\ell^2$, $k$, $M$ and branches complicatedly so that a variety of global structures for the solutions are found. 
In our analysis, the $\tilde{M}$-$r$ diagram is used, which
makes our consideration clear and enables easy understanding by visual effects.
In the plus branch,  all the solutions have the same  asymptotic structure at infinity as that in general relativity with a negative cosmological constant.
For the negative mass parameter, a new type of singularity called  the branch singularity appears at non-zero finite 
radius $r=r_b>0$. The divergent behavior around the singularity in Gauss-Bonnet gravity is milder than that around the central singularity in general relativity.
There are three types of horizons: an inner,  black hole, and  cosmological.
In the $k=1,~0$ cases the plus-branch solutions do not have any horizon.
In the $k=-1$ case, the radius of the horizon is restricted as $r_h<\sqrt{2\tilde{\alpha}}$ ($r_h>\sqrt{2\tilde{\alpha}}$) in the plus (minus)
branch. 
The black hole solution with zero or negative mass  exists in the
plus branch
even for the zero or positive cosmological constant.
There is also the extreme black hole solution with positive mass in spite of the
lack of electromagnetic charge. 
We briefly discuss the effect of the Gauss-Bonnet corrections on black hole formation in a collider and the possibility of the violation of third law of the black hole thermodynamics.
\end{abstract}




\maketitle

\section{Introduction}
\label{Intro}

Black holes are characteristic objects to general theory of relativity. 
Recent observational data show the existence of one  or more huge
black holes in the central region of a number of galaxies. 
While over the past decades much concerning the nature of  black
hole spacetimes has been clarified, a good many
unsolved problems remain. One of the most important ones is what the final
state of  black-hole evaporation through quantum effects is. 
The mid-galaxy supermassive black holes are certainly not related to this
problem; however, it has been suggested that tiny black holes, whose quantum
effect should not be neglected, could be formed in the early universe by the
gravitational collapse of the primordial density fluctuations.  Black holes may become
small enough in the final stage of evaporation enough for quantum 
aspects of gravity to
become noticeable. In other words, such tiny black holes may provide a good opportunity for learning not only about strong gravitational fields but also about of the quantum aspects
of gravity. 

Up to now many quantum theories of gravity have been proposed.
Among them superstring/M-theory formulated in the higher dimensional 
spacetime is the most promising candidate. 
So far, however, no much is known about
the non-perturbative aspects of the theory have not been
To
take string effects perturbatively into classical gravity is one 
approach to the study of  the quantum effects of gravity.  

A recent and attractive proposal for a
new picture of our universe called the braneworld
universe~\cite{large,Randall,DGP} is based on superstring/M-theory~\cite{Lukas}. 
According to these we live on a four-dimensional timelike hypersurface 
embedded in a
higher-dimensional bulk spacetime. 
Because the fundamental scale could be around TeV
scale in this scenario, these models suggest that creation of tiny black holes in
a linear hadron collider is possible~\cite{Bhformation}. From this point of
view, the investigation of the string effects on the black holes is important.

In this paper we consider the $n$-dimensional action with the Gauss-Bonnet terms for
gravity~\cite{GB}, which we call Gauss-Bonnet gravity hereafter, as the higher
curvature corrections to general relativity. The Gauss-Bonnet terms naturally arise as
the next leading order of the $\alpha'$-expansion of superstring theory, where $\alpha'$
is inverse string tension~\cite{Gross}, and are ghost-free combinations~\cite{Zwie}. 
The black hole solutions in Gauss-Bonnet gravity were first discovered by Boulware
and Deser~\cite{GB_BH} and Wheeler~\cite{Wheeler_1}, independently. 
Since then many types of black hole solutions have been intensively studied~\cite{Torii}.
The black hole
solutions with a cosmological constant were investigated in several
papers~\cite{Cho,Cvetic,Cai,Neupane,Cai2}. In the system with a negative
cosmological constant, black holes can have horizons with non-spherical
topology such as torus, hyperboloid, and other compactified sub-manifolds. These
solutions were originally found in general relativity and are called topological
black holes~\cite{Brown}.  They are distinctive in that there exist
zero-mass and negative mass black hole solutions, and in both cases their properties were
investigated~\cite{Cai3}.  Topological black hole solutions have also been studied in
Gauss-Bonnet gravity~\cite{Cho,Cvetic,Cai,Neupane}. 
Black hole solutions in more general Lovelock gravity~\cite{Lovelock} have also been investigated~\cite{Myers,Whitt}

In Gauss-Bonnet gravity, there are two kinds of black hole solutions, which are
classified into the plus and the minus branches.  When Boulware and Deser first
discovered the solutions, they claimed that the vacuum state in one of the branches (the
plus branch in our definition) are unstable~\cite{GB_BH}, so that the solutions in the minus branch have been
intensively investigated, while less attention has been paid to the solutions in the plus branch.  However, the vacuum states in both branches have recently turned out
to be stable~\cite{Deser}. Moreover, it is known that the solutions in
five-dimensional Gauss-Bonnet gravity have qualitatively different properties from those of
higher-dimensions.  There is, however, a lack of detailed investigations of the
five-dimensional case.

In this paper, we extend the previous work and  give a unified viewpoint on the black hole solutions in
Gauss-Bonnet gravity.  We include all the aforementioned cases,
i.e., the plus and the minus branches, five- and higher-dimensional cases.  In particular,
we focus on the global structure of the spacetime.  We also consider the case with a
positive/negative cosmological constant and that with a sub-manifold with non-spherical
topology.  We investigate not only the black hole solutions but also other kinds of
solutions, such as regular or globally naked solutions.

In Sec.~\ref{Model}, we introduce our model and show solutions  that are
generalizations of Boulware and Deser's original. In Sec.~\ref{EL}, we
review the solutions in the Einstein-$\Lambda$ system for comparison. 
In Sec.~\ref{EGBL}, the general properties of the solutions in the Einstein-Gauss-Bonnet-$\Lambda$
system with $4\tilde{\alpha}/\ell^2<1$, whose meaning is given in the text, are investigated. 
In Sec.~\ref{MR-diagram}, we show the $\tilde{M}$-$r$ diagram and study the number of horizons for each solution.
The global structures of the
solution are summarized in tables.
Sec.~\ref{EGBL2} is devoted to the analysis of the  special case where
$4\tilde{\alpha}/\ell^2=1$.
In Sec.~\ref{Conclusion}, we give conclusions and discuss related issues and future work.
Throughout this paper we use units such that $c=\hbar=k_B=1$. As for notation and
conversion we follow Ref.~\cite{Gravitation}. 
The Greek indices run $\mu$, $\nu=0,1, \cdots, n-1$.

\section{Model and solutions}
\label{Model}

We start with the following $n$-dimensional ($n\geq 4$) action:
\begin{equation}
\label{action}
S=\int d^nx\sqrt{-g}\biggl[ 
\frac{1}{2\kappa_n^2}
(R-2\Lambda+\alpha{L}_{GB}
) \biggr]
+S_{\rm matter},
\end{equation}
where
$R$ and $\Lambda$ are the $n$-dimensional 
Ricci scalar and
the cosmological constant, respectively. $\kappa_n:=\sqrt{8\pi G_n}$, where $G_n$ is the $n$-dimensional gravitational constant. The Gauss-Bonnet Lagrangean consists of the Ricci scalar,  the Ricci tensor, and the Riemann tensor as
\begin{equation}
{L}_{GB}=R^2-4R_{\mu\nu}R^{\mu\nu}
+R_{\mu\nu\rho\sigma}R^{\mu\nu\rho\sigma}.
\end{equation}
This set of terms is called the Gauss-Bonnet terms. 
In the four-dimensional case, the Gauss-Bonnet terms do not appear in the
equation of motion but contribute only as the surface terms. As a result, the model
becomes ordinary Einstein theory with a cosmological constant.
$\alpha$ is the coupling constant of the Gauss-Bonnet terms. This type of
action is derived from superstring theory in the low energy limit~\cite{Gross}.
In such cases
$\alpha$ is related to the inverse string tension and is positive definite.
Since the Minkowski
spacetime becomes unstable for negative $\alpha$, we consider only the
$\alpha\geq 0$ case in this paper. $S_{\rm matter}$ is the action of matter fields.

The gravitational equation of the action (\ref{action}) is
\begin{equation}
\label{g-eq}
{G}_{\mu\nu} +\alpha {H}_{\mu\nu} +\Lambda g_{\mu\nu}
= \kappa_n^2 
{T}_{\mu\nu},
\end{equation}
where 
\begin{eqnarray}
{G}_{\mu\nu}&:=&R_{\mu\nu}-{1\over 2}g_{\mu\nu}R,\\
{H}_{\mu\nu}&:=&2\Bigl[RR_{\mu\nu}-2R_{\mu\alpha}R^\alpha_{~\nu}
-2R^{\alpha\beta}R_{\mu\alpha\nu\beta}
\nonumber
\\
&& ~~~~
 +R_{\mu}^{~\alpha\beta\gamma}R_{\nu\alpha\beta\gamma}\Bigr]
-{1\over 2}g_{\mu\nu}{L}_{GB}.
\end{eqnarray}
Since we are interested in vacuum solutions in this paper, the energy-momentum tensor
$T_{\mu\nu}$ is set at zero.

We assume a static spacetime
and adopt the following line element:
\begin{equation}
\label{metric}
ds^2=-f(r)e^{-2\delta(r)}dt^2
+f^{-1}(r)dr^2+r^2d\Omega_{n-2}^2,
\end{equation}  
where $d\Omega_{n-2}^2=\gamma_{ij}dx^i dx^j$ 
is the metric of the $(n-2)$-dimensional Einstein space $M^{n-2}$. 

In the four-dimensional case, the Einstein space $M^2$ is two-dimensional.
By taking $M^2$ to be complete, we can write it as a quotient space
$M^2=\tilde{M}^2/\Gamma$,
where the universal covering space $\tilde{M}^2$ is either two-sphere $S^2$, 
two-torus 
$T^2$ or two-hyperboloid  $H^2$, and $\Gamma$ is a freely and properly discretely acting
subgroup of the isometry group of $\tilde{M}$. If the action of $\Gamma$ on
$\tilde{M}^2$ is nontrivial, ${M}^2$ is multiply connected. For concrete examples,
see Ref.~\cite{Wolf}.
In the five-dimensional case, i.e., the case in which the sub-manifold is the
three-dimensional Einstein space, the situation is almost the same~\cite{3-einstein_2}.
In the higher-dimensional cases, the $(n-2)$-dimensional ($n>5$) Einstein space 
has  rich structures and
is not necessarily homogeneous, i.e., it can even have non-constant
curvature~\cite{3-einstein_1}.

The $(t, t)$ 
and $(r, r)$ components of the gravitational equation (\ref{g-eq})
give $\delta'\equiv 0$, where the prime denotes a derivative with respect to $r$. By rescaling the time coordinate suitably, 
we can always set
\begin{equation}
\label{delta}
\delta \equiv 0,
\end{equation}  
without loss of generality.

The equation of the metric function $f$ is written as  
\begin{eqnarray}
&&rf'-(n-3)(k-f)-\frac{n-1}{\ell^2}r^{2}
\nonumber \\
&& ~~~~~
+\frac{\tilde{\alpha}}{r^2}(k-f)
\Bigl[2rf'-(n-5)(k-f)\Bigr]=0,
\end{eqnarray}  
where 
$\tilde{\alpha}:=(n-3)(n-4)\alpha$ and $\Lambda=-(n-1)(n-2)/(2\ell^2)$. In our
definition, a negative (positive) $\Lambda$ gives a positive (negative) $\ell^2$. $k$ is the curvature
of the $(n-2)$-dimensional Einstein space and takes $1$ (positive curvature), $0$ (flat),
and $-1$ (negative curvature).
This equation is integrated
to give the general solution as
\begin{equation}
\label{f-eq}
f
=k+\frac{r^2}{2\tilde{\alpha}}
\Biggl\{1
\mp\sqrt{1+4\tilde{\alpha}
\biggl(\frac{\tilde{M}}{r^{n-1}}-\frac{1}{\ell^2}\biggr)}\Biggr\},
\end{equation}  
where
\begin{eqnarray}
\tilde{M}:=\frac{16\pi G_n M}{(n-2)\Sigma_{n-2}^k}.
\end{eqnarray}  
The integration constant $M$ is proportional to the mass of a 
black hole for black hole spacetime. Although
we will also consider solutions which are not black holes, we call $M$
the mass of the solution.
$\Sigma_{n-2}^k$ is the volume of the unit $(n-2)$-dimensional Einstein
space. 
For example, $\Sigma_{n-2}^1=2\pi^{(n-1)/2}/\Gamma[(n-1)/2]$
for the $(n-2)$-dimensional sphere $(k=1)$, where $\Gamma$ is the gamma function.
There are two families of solutions that correspond to the sign in front of the
square root in Eq.~(\ref{f-eq}). We call the family with a minus (plus) sign
the minus- (plus-) branch solution.
By introducing new variables as $r:=r/\ell$, 
$\bar{M}:=\tilde{M}/\ell^{n-3}$, and
$\bar{\alpha}:=\tilde{\alpha}/\ell^2$,
the curvature radius $\ell$ is scaled out when the
cosmological constant is non-zero. In the $\Lambda=0$ case,
it is convenient to rescale the variables by $\tilde{\alpha}$
as $r:=r/\sqrt{\tilde{\alpha}}$ and
$\bar{M}:=\tilde{M}/\sqrt{\tilde{\alpha}}$.

The global structure of the spacetime is characterized by the properties of the 
singularities, horizons, and infinities. 
A horizon is a null hypersurface defined by $r=r_h$ such that $f(r_h)=0$
with finite curvatures, where $r_h$
is a constant horizon radius.  
In this paper, we call a horizon on which $df/dr|_{r=r_h}>0$ a black hole horizon~\cite{Penrose}.
If there is a horizon inside of a black hole horizon and if it satisfies 
$df/dr|_{r=r_h}<0$, we call it an inner horizon.
If a horizon satisfies $df/dr|_{r=r_h}<0$ and if it is the outermost
horizon, we call it a cosmological horizon.  We call a horizon on which
$df/dr|_{r=r_h}=0$ a degenerate horizon.  
Among them if the first non-zero derivative coefficient is positive, i.e.,
$d^pf/dr^p|_{r=r_h}>0$ and $d^qf/dr^q|_{r=r_h}=0$ for any $q<p$, $q, p \in {\rm N}$,
we call it a degenerate black hole horizon.
A solution with degenerate horizons is called extreme. 
The solutions in this paper are classified into three types by the existence of the
horizons. The first one is a black hole solution that has
a black hole horizon.  
A solution that does not have a black hole horizon but does have a locally naked singularity is a
globally naked solution.  This is the second type.
The last type is a singularity-free solution without a black hole horizon, which we call a regular solution. 

Since there are many parameters in our solutions, such as $\tilde{\alpha}$,
$\ell^2$  (or $\Lambda$), $\tilde{M}$, $k$ and $\pm$ branches, 
the analysis should be
performed systematically. In this paper we employ  a $\tilde{M}$-$r$ diagram, which
we explain below.

First we choose one of the branches, and  fix the values of $\Lambda$, 
$\tilde{\alpha}$, and $k$. Then the solutions are
parametrized by one parameter, $\tilde{M}$.
By changing the mass parameter $\tilde{M}$ of the
solutions, we can plot the location of the singularity
and horizons on the $\tilde{M}$-$r$ diagram.  
These plots form
a number of curves. The curve of the horizon radius may be a multi-valued
function of $\tilde{M}$, which means that there exist
several horizons in the spacetime, such as a black hole, an inner,
and a cosmological horizon. This diagram shows the number and location
of the horizons and singularities for each $\tilde{M}$.
In several cases the curve of the horizon radius becomes vertical
where two types of horizon coincide. The solution at any such vertical
point is extreme.
Next, when we vary other parameters, say $\tilde{\alpha}$, the curves on
the $\tilde{M}$-$r$ diagram slide a bit from the original one.
When there is an extreme solution
in the family of the solutions,
we obtain the curve for the
extreme solution on the $\tilde{M}$-$r$ diagram by joining points for extreme solutions 
in each $\tilde{\alpha}$. We will show concrete examples of the $\tilde{M}$-$r$ diagram in general relativity in the 
next section.

All the solutions at
vertical points of the $\tilde{M}$-$r_h$ curve are not extreme solutions. Let us make the relation clear between 
the vertical points and the extreme solutions.
When $\tilde{\alpha}$, $k$, and $\Lambda$ are fixed, the metric function $f$ is a function of
$r$ with one parameter $\tilde{M}$. Then let us define a new function 
$\bar{f}(r,\, \tilde {M}):=f(r)$, where the right-hand side (rhs) is evaluated with the 
mass parameter $\tilde {M}$.
The $\tilde{M}$-$r_h$ curve is obtained by the constraint condition $\bar{f}(r_h, \tilde{M})=0$.
Then along the $\tilde{M}$-$r_h$ curve, we find $d\tilde{M}/dr_h = -(\partial \bar{f}/\partial r)/(\partial \bar{f} / \partial \tilde{M})|_{r=r_h}$
for $\partial \bar{f} / \partial \tilde{M}\ne 0$. 
At the vertical point,  $d\tilde{M}/dr_h =0$. This implies that (i) $\partial \bar{f}/\partial r|_{r=r_h}=0$ and $\partial \bar{f}/\partial \tilde{M}|_{r=r_h}<\infty$,  or  
(ii) $\partial \bar{f}/\partial \tilde{M}|_{r=r_h}=\infty$.
The case  (i) exemplifies the degenerate horizon condition. In the latter case (ii), however,
one cannot be sure whether the horizon is degenerate or not. Then we should examine the
condition $f=d f/d r=0$ directly.
As we will see in the following sections, the latter case appears only when 
$r_h \to 0$ and $r_h \to r_b$, where $r_b$ is the radius of the branch singularity, 
defined in Sec.~\ref{EGBL}. The solution with $r_h=0$ or $r_h = r_b$
does not have a horizon  but, instead,  has a regular or a singular point at $r=r_h$.
As a result, only the case (i) produces a degenerate horizon.
In fact, however,  several solutions exist with $r_h \to 0$ and 
$d f/d r|_{r=r_h} \to 0$, both in general relativity and Gauss-Bonnet theory. We call this type of horizon an almost degenerate horizon.
These infinitesimally small black hole solutions may be important from the thermodynamical point of view because their temperature is also infinitesimally small and affects the evolution of the black hole.
If $\partial \bar{f} / \partial \tilde{M} = 0$, we find also $\partial \bar{f}/\partial r|_{r=r_h}=df/dr|_{r=r_h}=0$,
and the horizon is degenerate.
Details in general relativity and Gauss-Bonnet cases  will be discussed in each section.

\section{Static solutions in the Einstein-$\Lambda$ system}
\label{EL}

Before investigating  the system with the Gauss-Bonnet terms, we first
review the spacetime structure of the static
solutions described by the metric form (\ref{metric})  in the Einstein-$\Lambda$ system using the $\tilde{M}$-$r$ diagram for comparison.

By taking the limit of $\alpha\to 0$, Gauss-Bonnet gravity reduces to
general relativity. 
In this limit,
\begin{eqnarray}
\label{einstein}
f=k-\frac{\tilde{M}}{r^{n-3}}+\frac{r^2}{\ell^2}
\end{eqnarray}  
is obtained only for the minus branch of Eq.~(\ref{f-eq}), while there is no such
limit for the plus branch.
The spacetime is described by Eqs.~(\ref{metric}), 
(\ref{delta}), and (\ref{einstein}).
We do not consider the case with $\tilde{M}=k=\Lambda= 0$, in which the
function $f$ is identically zero.

\begin{figure}
\includegraphics[width=.95\linewidth]{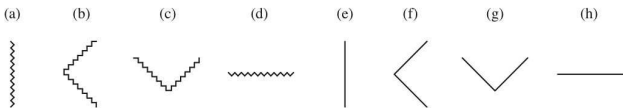}
\caption{
The conformal diagrams around a regular center and a singularity. Thick and wavy lines denote a regular center and a singularity, respectively. 
There are time-reversed diagrams for (c), (d), (g), and (h).
}
\label{penrose-center}
\end{figure}

There is a curvature singularity  at the center ($r=0$) except for the case with
$\tilde{M}=0$.  Around the center, the Kretschmann invariant behaves as follows:
\begin{eqnarray}
{\cal I}&:=&R_{\mu\nu\rho\sigma}R^{\mu\nu\rho\sigma}
\nonumber
\\
&=&
(f'')^2+\frac{2(n-2)}{r^2}(f')^2
+\frac{2(n-2)(n-3)}{r^4}(k-f)^2
\nonumber 
\\
&=& 
O\Bigl(\frac{\tilde{M}^2}{r^{2n-2}}\Bigr).
\label{kretchemann_e}
\end{eqnarray}  
Whether the central singularity is spacelike, null, or timelike
depends on the dominant term of Eq.~(\ref{einstein}) for $r \to 0$.
When $\tilde{M}> 0$ ($\tilde{M}< 0$), the metric function $f<0$ ($f>0$),
and the tortoise coordinate defined by 
\begin{equation}
r^{\ast}:=\int^{r} f^{-1}dr,
\label{tortoise}
\end{equation}  
is finite at the center. 
So the singularity is spacelike (Fig.~\ref{penrose-center}(d))
(timelike (Fig.~\ref{penrose-center}(a))).  
When $\tilde{M}=0$ and $k=1$ ($k=-1$), the center $r=0$ is 
regular~\cite{regular} and  timelike (Fig.~\ref{penrose-center}(e))
(spacelike (Fig.~\ref{penrose-center}(h))), while it becomes null for
$k=0$ and $\ell^2\ne 0$  because $|r^{\ast}| \to \infty$
(Fig.~\ref{penrose-center}(f) for $\ell^2>0$ 
and Fig.~\ref{penrose-center}(g) for $\ell^2< 0$).

\begin{figure}
\includegraphics[width=.52\linewidth]{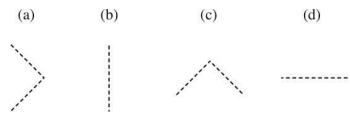}
\caption{
The conformal diagrams of the region of $r \to \infty$.
There are time-reversed diagrams for (c) and (d).
}
\label{penrose-infty}
\end{figure}


The structure of infinity  depends on the dominant term of Eq.~(\ref{einstein}) for 
$r \to \infty$. 
When there is a non-zero cosmological constant, the dominant term is 
$r^2/\ell^2$. 
When $1/\ell^2<0$ ($1/\ell^2>0$), the spacetime asymptotically
approaches  the de Sitter (dS) (anti-de Sitter (adS)) spacetime, of which region is expressed by the
conformal diagram Fig.~\ref{penrose-infty}(d)
(Fig.~\ref{penrose-infty}(b)). The second leading term is the
curvature term $k$. 
If $1/\ell^2=0$ and $k=1$, the spacetime is asymptotically flat, of which region is expressed by the conformal diagram Fig.~\ref{penrose-infty}(i).
If $1/\ell^2=0$ and $k=-1$, the conformal diagram of infinity is Fig.~\ref{penrose-infty}(iii).
When $1/\ell^2=k=0$, the $\tilde{M}$ term is dominant.
For $\tilde{M}>0$ ($\tilde{M}<0$), infinity is expressed by Fig.~\ref{penrose-infty}(iii) (Fig.~\ref{penrose-infty}(i)).


Since the number of the horizons depends complicatedly on parameters such as  mass,  
the cosmological constant, and so on, the $\tilde{M}$-$r$ diagram is useful.
From Eq.~(\ref{einstein}), the $\tilde{M}$-$r_h$ relation in the
Einstein-$\Lambda$ system is 
\begin{eqnarray}
\label{M-horizon-E}
\tilde{M}=r_h^{n-1}\biggl(\frac{1}{\ell^2}
+\frac{k}{r_h^2}
\biggr).
\end{eqnarray}  
We consider here that the values of $k$ and $\ell^2$ are
fixed, so that this relation becomes a curve on the $\tilde{M}$-$r$
diagram.

On the $\tilde{M}$-$r_h$ curve there are some points that are physically important.
For example,  points where the curve terminates at a singularity,
points where the value of $r_h$ becomes zero, and  points where the curve becomes
vertical. 
From the degeneracy condition of the horizon $f(r_{ex})=df/dr|_{r=r_{ex}}=0$, where $r_{ex}$ is the radius of the degenerate horizon, we can show that the $\tilde{M}$-$r_h$ curve becomes vertical at this point. 
However, the inverse is not always true, as we showed in Sec.~\ref{Model}.
Let us examine the conditions (i) and (ii) in Sec.~\ref{Model} in general relativity.
By Eq.~(\ref{einstein}), $\partial \bar{f}/ \partial \tilde{M}|_{r=r_{h}} = -r_h^{-(n-3)}$. Except for $r_h\to 0$, this equation becomes finite. Hence if the horizon radius at the vertical point on the $\tilde{M}$-$r_h$ curve is non-zero, the horizon is degenerate.
In the $r_h\to 0$ case,  $\tilde{M}$ becomes zero by Eq.~(\ref{M-horizon-E}). If we set $\tilde{M}=0$ exactly,
the physical center is not a horizon but a regular point. 
However,  the analysis of the limiting solution $r_h\to 0$, i.e., 
an infinitesimally small horizon, is important when we consider the evolution of the 
black hole.
The condition of the degenerate horizon $f=df/dr=0$ gives
\begin{eqnarray}
\label{r-extreme-E}
\frac{df}{dr}\biggl|_{r=r_h}
= \frac{(n-3)k}{r_h}+\frac{(n-1)r_h}{\ell^2}.
\end{eqnarray}  
In the $r_h\to 0$ case, $df/dr|_{r=r_h}=0$ for $k=0$ and $df/dr|_{r=r_h}\to \infty$ for $k=\pm 1$.
Hence for $k=0$ the horizon of the vertical point with $r_h\to 0$ is the almost degenerate horizon, while it is not for  $k=\pm 1$.

The location and  the mass parameter of the degenerate horizon are 
\begin{eqnarray}
\label{r-extreme-E}
&&r_{ex}=\sqrt{-\frac{(n-3)k\ell^2}{n-1}},
\\
\label{M-extreme-E}
&&\tilde{M}_{ex}
=\frac{2kr_{ex}^{n-3}}{n-1},
\end{eqnarray}  
respectively, for $\Lambda\ne 0$.
The extreme
solution exists when $\ell^2$ and $k$ have signs  opposite to each
other. When $k= 0$, these equations give $\tilde{M}_{ex}= r_{ex}=0$ and the solution is not  an extreme solution but a regular solution.
For $\Lambda = 0$, the degeneracy condition gives $\tilde{M}= k=0$. This solution
falls outside of our consideration.
If $d^2\tilde{M}/dr_h^{~2}=0$ is satisfied at the finite value of $r_h$,
the three horizons degenerate. However, there is no such solution in the present system.

\begin{widetext}

\begin{figure}
\includegraphics[width=.70\linewidth]{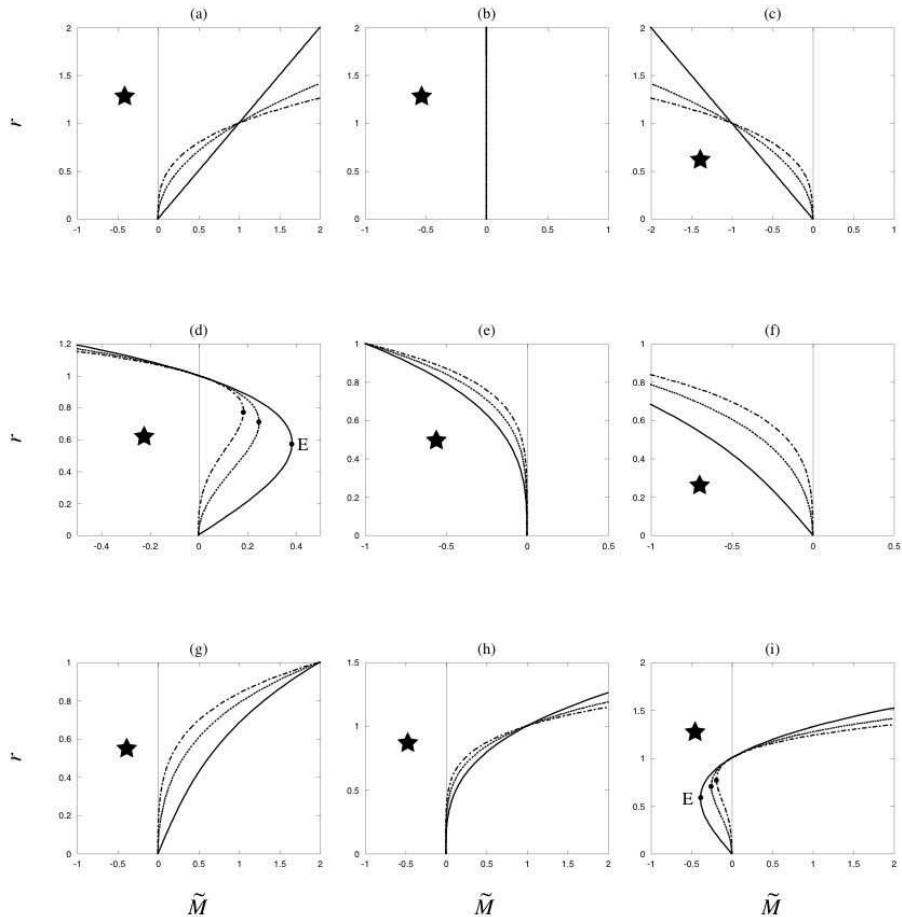}
\caption{
The $\tilde{M}$-$r$ diagrams for the static
solutions in the
Einstein-$\Lambda$ system. 
The diagrams in the upper, middle, and lower rows are 
$1/\ell^2=0$ (zero cosmological constant), 
the $1/\ell^2=-1$ (positive cosmological constant), and
the $1/\ell^2=1$ (negative cosmological constant) cases,
respectively.
The diagrams in the left, middle, and right
columns are the $k=1$, $0$, and $-1$ cases, respectively.
The $\tilde{M}$-$r_h$ relations in $n=4$ (thick solid curve), 
in $n=5$ (thick dashed curve), and in $n=6$ (thick dot-dashed curve) are shown.
There is a central singularity at $r=0$ except for $\tilde{M}=0$.
The dots with character ``E"  imply the degenerate horizons.
The region with a star which is bounded by the $\tilde{M}$-$r_h$
curve is an untrapped region.
It changes from  untrapped to  trapped region or vice versa by crossing
the $\tilde{M}$-$r_h$ curve.
}
\label{fig:E_m-rh}
\end{figure}

\end{widetext}

On the $\tilde{M}$-$r$ diagram the extreme solution is expressed by the point where
two horizons coincide. 
The diagram in Fig.~\ref{fig:E_m-rh} (d) is a typical example.
This solution is the Schwarzschild-dS solution. The point
E is where the curve becomes vertical. 
The solution at this point is the extreme solution and has the mass
parameter $\tilde{M}_{ex}$.  A solution with less mass $\tilde{M}<\tilde{M}_{ex}$ has two
horizons (a black hole and a cosmological horizons), and at the extreme
point two horizons coincide. For the greater mass parameter, there is no
horizon. In this manner, the number of the horizons changes at the mass parameter of the
extreme solution like this.
The number of the horizons also changes with the mass parameter at the points where
the curve terminates in singularities, 
the value of $r_h$ becomes zero  for a finite $\tilde{M}$, and so on.
We can see these points directly on the $\tilde{M}$-$r$ diagram.


We will examine the number of horizons of the solutions
by varying the mass parameter and show their spacetime
structures in the cases of zero, positive,
and negative cosmological constant, separately.
The $\tilde{M}$-$r$ diagrams are shown in Fig.~\ref{fig:E_m-rh}
and the global structures of the solutions are summarized in
Table~\ref{E-penrose}.

\subsection{$\Lambda=0$ case}

In the $k=1$ case, 
the solution is the $n$-dimensional Schwarzschild solution.
The $\tilde{M}$-$r$ diagram is shown in Fig.~\ref{fig:E_m-rh}(a).
For $\tilde{M} < 0$, the solution has no horizon and represents the spacetime with a
globally naked singularity. 
The solution with $\tilde{M}=0$ is the Minkowski spacetime~\cite{note_1}. 
For $\tilde{M} > 0$, the solution has a single horizon at 
$r_h=\tilde{M}^{1/(n-3)}$ and represents
the Schwarzschild black hole spacetime.

In the $k=0$ case, the $\tilde{M}$-$r$ diagram is shown in Fig.~\ref{fig:E_m-rh}(b).
The $\tilde{M}$-$r_h$  relation becomes $\tilde{M}=0$. However, since we
do not consider the $\tilde{M}=k=\Lambda=0$ case, there is no
$\tilde{M}$-$r_h$ curve in this diagram. 
For $\tilde{M}\ne 0$, the solution has no horizon and represents the spacetime with a
globally naked singularity.

In the $k=-1$ case, the $\tilde{M}$-$r_h$ diagram is shown in Fig.~\ref{fig:E_m-rh}(c). 
For $\tilde{M} < 0$, the solution has a cosmological horizon and represents the spacetime with a
globally naked singularity. 
For $\tilde{M} = 0$, the solution is
regular hyperbolic spacetime. 
For $\tilde{M}> 0$, the solution  has no
horizon and represents the spacetime with a
globally naked singularity.

\subsection{$\Lambda>0$ case}

In the $k=1$ case, the solution is the Schwarzschild-dS solution.
The $\tilde{M}$-$r_h$ diagram is shown in Fig.~\ref{fig:E_m-rh}(d). 
There is an extreme solution with mass $\tilde{M}_{ex}$ given by
Eq.~(\ref{M-extreme-E}).
For $\tilde{M}<0$, the solution has a
cosmological horizon and represents the spacetime with a
globally naked singularity.
For $\tilde{M}=0$, the spacetime is the $n$-dimensional dS spacetime,
which has a cosmological horizon at $r=\sqrt{|\ell^2|}$.
For $0<\tilde{M}<\tilde{M}_{ex}$,
there are a black hole and a cosmological horizons and the solution represents the Schwarzschild-dS black hole spacetime.
For $\tilde{M}=\tilde{M}_{ex}$, the black hole and the cosmological
horizons coincide, and the horizon is degenerate. The location of the
degenerate horizon is given by Eq.~(\ref{r-extreme-E}). 
For $\tilde{M}>\tilde{M}_{ex}$, the solution has no horizon and represents the spacetime with a
globally naked singularity.

In the $k=0$ case, the $\tilde{M}$-$r_h$ diagram is shown in 
Fig.~\ref{fig:E_m-rh}(e). 
For $\tilde{M}<0$, the solution has a cosmological horizon and represents the spacetime with a
globally naked singularity. The derivative of the
metric function vanishes in the zero horizon limit, i.e.,
$f'(r_h)\to 0$ as $r_h\to 0$. This implies that the horizon 
almost degenerates in this limit.
For $\tilde{M}=0$, the solution has no horizon and represents the
regular spacetime. 
For $\tilde{M}>0$, the solution has no horizon and represents the spacetime with a
globally naked singularity.

In the $k=-1$ case, the $\tilde{M}$-$r_h$ diagram is shown in 
Fig.~\ref{fig:E_m-rh}(f). 
For $\tilde{M}<0$, the solution has a cosmological horizon and represents the spacetime with a
globally naked singularity. 
For $\tilde{M}=0$, the solution has no horizon and represents the
regular spacetime. 
For $\tilde{M}>0$, the solution has no horizon and represents the spacetime with a
globally naked singularity.

\subsection{$\Lambda<0$ case}

When the curvature of the
$(n-2)$-dimensional Einstein space is $k=0$ or $-1$, there is no black hole solution in the zero or positive cosmological constant case.
However, the negative cosmological constant allows such
black hole solutions. They are called topological black
holes~\cite{Brown,Cai3} and show remarkable properties in that the mass of such
solutions can be negative.

In the $k=1$ case, the solution is  the $n$-dimensional Schwarzschild-adS solution.
The $\tilde{M}$-$r_h$ diagram is shown in 
Fig.~\ref{fig:E_m-rh}(g).
For $\tilde{M}<0$, the solution has no horizon and represents the spacetime with a
globally naked singularity. 
For $\tilde{M}=0$, the solution has no horizon and represents the everywhere regular spacetime.
The spacetime is the adS spacetime. 
For $\tilde{M}>0$, the solution has one black hole horizon and represents the Schwarzschild-adS black hole spacetime.

In the $k=0$ case, the $\tilde{M}$-$r_h$ diagram is shown in Fig.~\ref{fig:E_m-rh}(h).
For $\tilde{M}<0$, the solution has no horizon and represents the spacetime with a
globally naked singularity. 
For $\tilde{M}=0$, the solution has no horizon; the
spacetime has a regular null center. 
For $\tilde{M}>0$, the solution has one black hole horizon. 
In the infinitesimally small horizon limit $r_h \to 0$, the horizon is almost degenerate.

Where $k=-1$, the case is quite different from the others.  
The black hole solution with zero or negative mass can exist.
The $\tilde{M}$-$r_h$ diagram is shown in Fig.~\ref{fig:E_m-rh}(i).
There is an extreme solution which has the lowest mass of
all the black hole solutions in this system:
\begin{eqnarray}
\tilde{M}_{ex}
=-\frac{2}{n-1}\biggl(\frac{n-3}{n-1}\ell^2\biggr)^{\frac{n-3}{2}}.
\end{eqnarray}  
The degenerate horizon locates at
\begin{eqnarray}
r= r_{ex}=\sqrt{\frac{n-3}{n-1}}\ell.
\end{eqnarray}  
For $\tilde{M}<\tilde{M}_{ex}$, the solution has no horizon and represents the spacetime with a
globally naked singularity.
For $\tilde{M}=\tilde{M}_{ex}$, the solution has a degenerate horizon and represents the extreme black hole spacetime.
For $\tilde{M}_{ex}<\tilde{M}<0$, the solution has 
an inner and a black hole horizons and represents the black hole spacetime with negative mass.
For $\tilde{M}=0$, the solution has regular
center and a black hole horizon at $r_h=l$. 
The spacetime is black hole spacetime with zero mass.
It should be noted again that the term ``regular" in this paper means that 
the Kretschmann invariant does not diverge. So the ``regular" point can be singular by the topological origin
of the $(n-2)$-dimensional Einstein space. The center of the zero mass black hole is this case.
For $\tilde{M}>0$, the solution has a black hole horizon and represents the black hole spacetime.

\section{General properties of solutions in the Einstein-Gauss-Bonnet-$\Lambda$
system
}
\label{EGBL}

Next we proceed to the cases of Gauss-Bonnet gravity. 
Let us discuss the general properties of the static solutions in this
system.


In the case of $\tilde{M}=0$,
the metric function is
\begin{eqnarray}
\label{metric-f1}
f=k+\frac{r^2}{2\tilde{\alpha}}
\biggl(1\mp\sqrt{1-\frac{4\tilde{\alpha}}{\ell^2}}\biggr).
\end{eqnarray}  
For a well-defined theory, the following condition should be satisfied:
\begin{eqnarray}
\label{vacuum}
\frac{4\tilde{\alpha}}{\ell^2} \leq 1.
\end{eqnarray}  
Although any non-negative value of $\tilde{\alpha}$ satisfies this condition in
the zero and positive cosmological constant cases, it is restricted as $0\leq\tilde{\alpha}\leq \ell^2/4$
where the cosmological constant is negative.
In this section, we assume $4\tilde{\alpha}/\ell^2<1$.
The special case where $4\tilde{\alpha}/\ell^2=1$ will be investigated in
Sec.~\ref{EGBL2} separately.


From Eq.~(\ref{metric-f1}), the effective curvature radius is
defined by
\begin{eqnarray}
\ell_{\rm eff}^2
:=\frac{\ell^2}{2}
\biggl(1\pm\sqrt{1-\frac{4\tilde{\alpha}}{\ell^2}}\biggr).
\end{eqnarray}  
The sign of the square of the effective curvature radius $\ell_{\rm eff}^2$ is 
the same as $\ell^2$ in the minus branch, while  $\ell_{\rm eff}^2$ is 
always positive independent of $\ell^2$ in the plus branch.
Hence the spacetime approaches the adS spacetime for $r \to \infty$
asymptotically in the plus branch for $k=1$ even when the pure cosmological 
constant $\Lambda$ is positive. 
In Sec.~\ref{EL}, we showed the structure of the infinity of the solutions,
which in the Einstein-$\Lambda$ system depends on the value of $\ell^2$.
By replacing $\ell^2$  in the discussion in the Einstein-$\Lambda$ system  by $\ell_{\rm eff}^2$,
the structure of infinity in the 
Einstein-Gauss-Bonnet-$\Lambda$ system is obtained.


Besides the central singularity at $r=0$, there can be another singularity
at $r=r_b$ called the branch singularity. $r_b$ is obtained by the condition
that the inside
of the square root of Eq.~(\ref{f-eq}) vanishes, and the $\tilde{M}$-$r_b$
relation becomes
\begin{equation}
\label{branch-sing}
\tilde{M}=\tilde{M}_b
:=-\Bigl(1-\frac{4\tilde{\alpha}}{\ell^2}\Bigr)
\frac{r_b^{n-1}}{4\tilde{\alpha}}.
\end{equation}  
Around the branch singularity the Kretschmann invariant
behaves as
\begin{equation}
\label{divbranch}
{\cal I}\sim O\bigl[(r-r_b)^{-3}\bigr].
\end{equation}  
By the condition Eq.~(\ref{vacuum}), there are no positive real roots
when $\tilde{M}\geq 0$, while one positive root exists for $\tilde{M}< 0$.
$r_b$ decreases monotonically as
$\tilde{M}$ increases and approaches  zero in the limit of
$\tilde{M} \to 0$.  

When a branch singularity exists, the metric function $f$ behaves around it as
\begin{equation}
f(r) \approx \biggl(k+\frac{r_b^2}{2\tilde{\alpha}}\biggl)  
\mp
\frac{r_b^2}{2\tilde{\alpha}}\sqrt{\frac{n-1}{r_b}\biggl(1-\frac{4{\tilde
\alpha}}{\ell^2}\biggl)} 
(r-r_b)^{1/2}. 
\end{equation}  
Hence if $k=1$ or $0$, the singularity is timelike (Fig.~\ref{penrose-center}(a)).
When $k=-1$, the singularities are timelike  
and spacelike (Fig.~\ref{penrose-center}(d)) for
$r_b>\sqrt{2\tilde{\alpha}}$ and
$r_b<\sqrt{2\tilde{\alpha}}$, respectively.
In the special case where $k=-1$ and $r_b=\sqrt{2\tilde{\alpha}}$, the
singularities are spacelike and timelike  for the minus and plus branches,
respectively.

The positive-mass solutions have a central singularity.
Here the metric function behaves
\begin{equation}
f \approx  \mp \sqrt{\frac{\tilde{M}}{\tilde{\alpha}r^{n-5}}}
+k+\frac{r^2}{2\tilde{\alpha}}
\mp\sqrt{\frac{1}{64\tilde{\alpha}^3\tilde{M}}}
\biggl(1-\frac{4\tilde{\alpha}}{\ell^2}\biggr)
r^{\frac{n+3}{2}}
\end{equation}  
around the center. Hence if $n\geq 6$, the singularity is 
spacelike  (timelike) for the minus (plus) branch.
The Kretschmann invariant behaves as
\begin{eqnarray}
\label{divGB}
{\cal I}\sim
O\Bigl(\frac{\tilde{M}}{r^{n-1}}\Bigr).
\end{eqnarray}  
Although $f$ is finite when $n=5$, the Kretschmann
invariant behaves as
\begin{eqnarray}
{\cal I}
\sim
\frac{12\tilde{M}}{\tilde{\alpha}r^4}+O(r^{-2}),
\end{eqnarray}  
so that the center is singular. 
In the minus branch, the singularity is spacelike
when $k=0$ and
$-1$. When $k=1$, the singularity is spacelike for 
$\tilde{M}> \tilde{\alpha}$, null
(Fig.~\ref{penrose-center}(b)) for $\tilde{M}=
\tilde{\alpha}$, and timelike 
for $0<\tilde{M}< \tilde{\alpha}$. In the plus branch, the
singularity is timelike when
$k=0$ and $1$. When $k=-1$, the singularity is spacelike for 
$0<\tilde{M}< \tilde{\alpha}$, null for $\tilde{M}=
\tilde{\alpha}$, and timelike 
for $\tilde{M}> \tilde{\alpha}$. It is noted that the
divergent behavior of the central singularity in
Gauss-Bonnet gravity   is milder than that in general
relativity (see Eqs.~(\ref{kretchemann_e}) and
(\ref{divGB})). It is also noted that the divergent behavior
of the branch singularity is milder than that of the central
singularity (see Eqs.~(\ref{divbranch}) and (\ref{divGB})).

For the solution with zero mass, the center is regular. 
Since the metric function is described as
$f=k+r^2/\ell_{\rm eff}^2$, the conformal diagram around the
center is obtained by replacing $\ell^2$ in the discussion in
the Einstein-$\Lambda$ system by $\ell_{\rm eff}^2$.


From the condition of the horizon $f(r_h)=0$, we find 
\begin{equation}
\pm\Bigl(1+\frac{2\tilde{\alpha}k}{r_h^2}\Bigr)
=\sqrt{1+4\tilde{\alpha}
\biggl(\frac{\tilde{M}}{r_h^{n-1}}-\frac{1}{\ell^2}
\biggr)}
>0.
\end{equation}  
Hence
\begin{eqnarray}
&& r_h^2<-2\tilde{\alpha}k, ~~~ (\mbox{\rm for the plus branch}),
\\
&& r_h^2>-2\tilde{\alpha}k, ~~~ (\mbox{\rm for the minus branch}),
\end{eqnarray}  
so that we can conclude that solutions in the plus branch have
no horizon for $k=1, 0$. 
Hence the solutions represent the spacetime with a
globally naked singularity  except for the solution with $\tilde{M}=0$.
When
$k=-1$, the horizon radius is restricted as
$r_h<\sqrt{2\tilde{\alpha}}$ in the plus branch, while 
in the minus branch, the horizon has the lower limit
$r_h>\sqrt{2\tilde{\alpha}}$. 
There are no such restrictions for $k=1$ and $0$.

The $\tilde{M}$-$r_h$ relation is
\begin{eqnarray}
\label{M-horizon}
\tilde{M}=r_h^{n-1}\biggl[\frac{1}{\ell^2}
+\frac{k}{r_h^2}\Bigl(1+\frac{\tilde{\alpha} k}{r_h^2}\Bigr)
\biggr].
\end{eqnarray}  
For $k=0$ this relation is same as that in general relativity Eq.~(\ref{M-horizon-E}).
For $k=-1$ the solution with $r_h=\sqrt{2\tilde{\alpha}}$
has a branch singularity at $r=r_B:=\sqrt{2\tilde{\alpha}}$,
where
\begin{equation}
\label{M-branch}
\tilde{M}=\tilde{M}_B
:=-\frac{(2\tilde{\alpha})^{\frac{n-3}{2}}}{2}
\Bigl(1-\frac{4\tilde{\alpha}}{\ell^2}\Bigr).
\end{equation}  
This implies that the sequence of solutions
is divided by the branch singularity into the plus and the minus
branches for $k=-1$. 
It is seen that the $\tilde{M}$-$r_h$ curve in the $\tilde{M}$-$r$ diagram 
terminates at the $\tilde{M}$-$r_b$ curve just at the point with  $\tilde{M}=\tilde{M}_B$ (Point B
in Figs.~\ref{fig:M-r-6-neutral} and \ref{fig:M-r-5-neutral}). 
For the other mass parameter $\tilde{M}\ne\tilde{M}_b$, the
horizon radius is always larger than that of
the branch singularity.
In the limit of $r_h\to 0$, the mass of the solution
vanishes
$\tilde{M}\to 0$ in the
$n\geq 6$ case, while it approaches
$\tilde{M}\to\tilde{\alpha}k^2$ in five-dimensions.


In general relativity there are some extreme black hole solutions
with a degenerate horizon. The Gauss-Bonnet case has more variety.
First let us examine the relation of a vertical point in the $\tilde{M}$-$r_h$ curve and
an extreme solution. By Eq.~(\ref{f-eq}), $\partial \bar{f}/\partial \tilde{M} |_{r=r_h}<\infty$ except for $r_h \to 0$ and $r_h \to r_b$. Hence the horizon of the vertical point is degenerate for  the $r_h \ne 0$ and $r_h \ne r_b$ cases.
In the $r_h \to 0$ case, $df/dr|_{r=r_h} \to 0$ for $k=0$ in $n\geq 6$ and for any $k$ in $n=5$, and the horizon is
almost degenerate, 
while $df/dr|_{r=r_h} \to \infty$ for $k=\pm 1$ in $n\geq 6$, and the horizon is not degenerate.
In the  $r_h \to r_b$ case, $df/dr|_{r=r_h} \to \infty$ for any $k$ and the  horizon is not degenerate.

The radius of the degenerate horizon
is
\begin{equation}
\label{extreme-neutral-1}
r_{ex}^2=\frac{(n-3)\ell^2}{2(n-1)}\biggl[-k\pm |k|
\sqrt{1-\frac{4\tilde{\alpha}}{\ell^2}\frac{(n-1)(n-5)}{(n-3)^2}}\biggr]
\end{equation}  
for $\Lambda \ne 0$,
and 
\begin{equation}
\label{extreme-neutral-2}
r_{ex}=\sqrt{\frac{-k(n-5)\tilde{\alpha}}{n-3}}
\end{equation}  
for  $\Lambda = 0$. Note that the sign $\pm$ does not correspond
to the sign of the solution branches.
The mass of the extreme solution is
\begin{equation}
\label{M-extreme}
\tilde{M}=\tilde{M}_{ex}=\frac{2k}{n-1}
\biggl(1+\frac{2\tilde{\alpha}k}{r_{ex}^2}\biggr)
r_{ex}^{n-3}.
\end{equation}  
We denote the mass of the extreme solution and the radius of  its degenerate horizon in the plus (minus) branch as $\tilde{M}_{ex}^{(+)}$ and  $r_{ex}^{(+)}$ 
($\tilde{M}_{ex}^{(-)}$ and  $r_{ex}^{(-)}$), respectively.
In the $n\geq 6$ and $\Lambda= 0$ case, $k$ must be
negative by the positivity of the degenerate horizon.
Since $r_{ex}<\sqrt{\tilde{\alpha}}$, the extreme solution
belongs to the plus branch. 
In the $\Lambda> 0$ case, the minus
sign should be taken in Eq.~(\ref{extreme-neutral-1}), and
the extreme solutions appear in both $k=1$ and $-1$ cases.
In the $\Lambda< 0$ case, $k$ must be negative, and then both signs
in Eq.~(\ref{extreme-neutral-1}) give extreme solutions.
Although the black hole solution with a
degenerate horizon necessarily has negative mass in general relativity (in the
$\Lambda<0$ and $k=-1$ case),
the extreme black hole
solutions  in the $k=-1$ cases   have positive mass.

In five-dimension the rhs of  Eq.~(\ref{extreme-neutral-2})
vanishes, and there is no extreme solution in the $\Lambda= 0$
case.  For $\Lambda> 0$ ($\Lambda< 0$), $k$ must be positive
(negative) and the minus (plus) sign should be assumed in 
Eq.~(\ref{extreme-neutral-1}).  
The radius of the degenerate horizon does not depend on $\tilde{\alpha}$,
and
Eq.~(\ref{extreme-neutral-1}) gives the same
expression as that in the Einstein-$\Lambda$ system. This implies that a degenerate horizon appears at
the same radius as that in the Einstein-$\Lambda$ system. However, the
mass of the extreme solution depends on $\tilde{\alpha}$, and the 
$\tilde{M}$-$r_{ex}$ relation shows a line on the $\tilde{M}$-$r$
diagram, as we will see in the next section.


The topological black hole solutions in general relativity can have a zero or
a negative mass parameter. Since the minus-branch solutions approach
the solutions in general relativity in the $\tilde{\alpha}\to 0$ limit,
it is expected that 
such solutions with  zero or
negative mass exist also in the Gauss-Bonnet case. 
Furthermore, similar solutions appear in the plus branch.
When the cosmological constant is zero, the horizon radius $r_0$ of the massless solution is
\begin{equation}
r_0^2=-\tilde{\alpha} k,
\end{equation}  
from Eq.~(\ref{M-horizon}).
There is one positive real root $\sqrt{\tilde{\alpha}}$ 
in the plus branch
for $k=-1$. This is a black hole horizon.
In the $\Lambda\ne 0$ cases, the horizon radius $r_0$ of the massless solution
is given as
\begin{equation}
\label{neutral-ex}
r_0^2=\frac{\ell^2}{2}\biggl(-k\pm |k|
\sqrt{1-\frac{4\tilde{\alpha}}{\ell^2}}\biggr) .
\end{equation}  
The sign $\pm$ does not mean the branches.
When $\Lambda<0$, $r_0$ has two positive real roots for $k=-1$,
one of which is in the plus branch while the other is in the minus branch, while
there are not such solutions for  $k=0$ and $1$.  Both horizons are black hole horizons. When
$\Lambda>0$, $r_0$ has one positive real root in the minus branch 
for $k=1$ and in the plus branch for $k=-1$. The former corresponds to
the conventional cosmological horizon in dS spacetime. The latter is, however,  the black hole
solution, as we will see later. Hence there are zero mass black hole solutions
even in the zero and positive cosmological constant cases. 
This type of solution does not
exist in general relativity with $\Lambda\geq 0$ but appears if the 
Gauss-Bonnet effects are considered.

\section{$\tilde{M}$-$r$ diagram and spacetime structure
}
\label{MR-diagram}

\begin{widetext}

\begin{figure}[tbp]
\includegraphics[width=.90\linewidth]{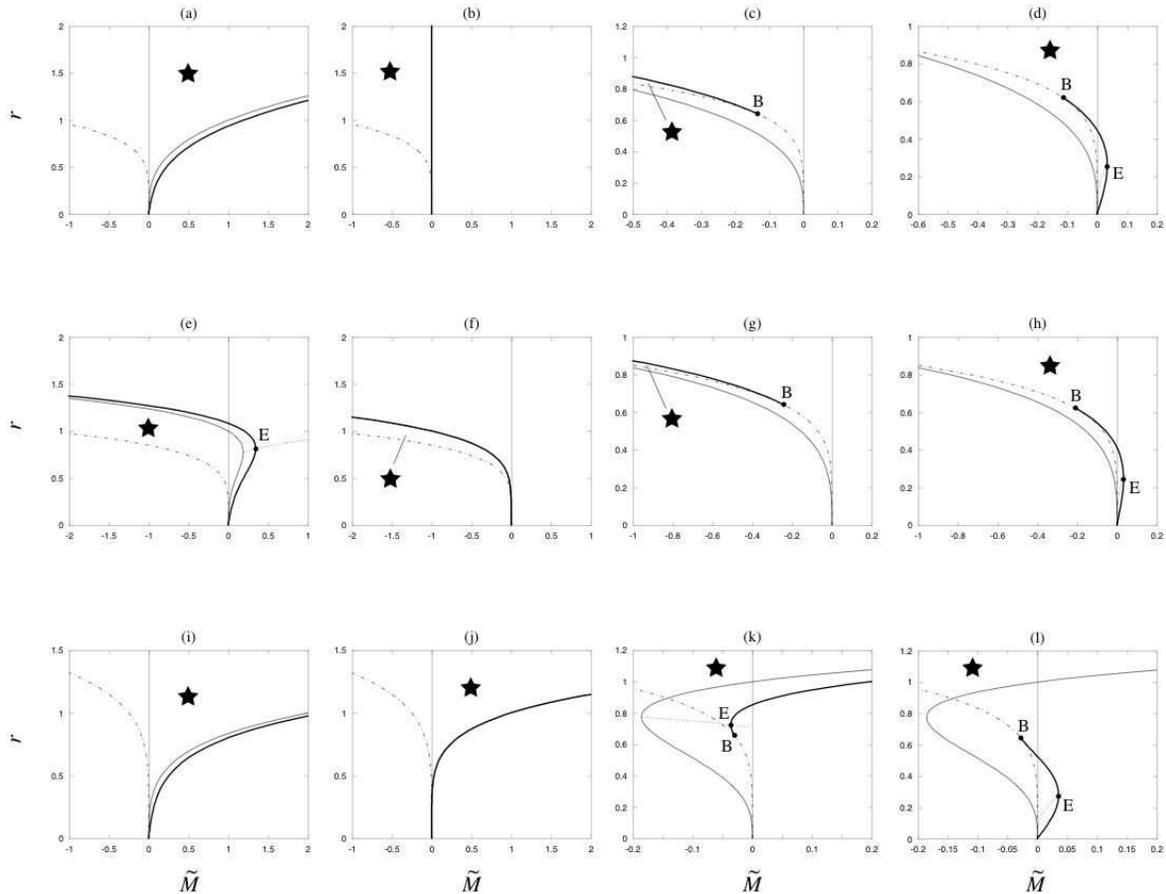}
\caption{
The $\tilde{M}$-$r$ diagrams of the static
solutions in the
six-dimensional Einstein-Gauss-Bonnet-$\Lambda$ system. 
The diagrams in the upper, the middle, and the lower rows express cases the 
$1/\ell^2=0$ (zero cosmological constant), 
the $1/\ell^2=-1$ (positive cosmological constant), and
the $1/\ell^2=1$ (negative cosmological constant),
respectively.
The diagrams in the columns on the left-hand side
express the $k=1$ (the minus branch), $0$ (the minus branch), $-1$  (the
minus branch), and  $-1$  (the plus branch) cases, respectively. 
We show
the $\tilde{M}$-$r_h$ relations (thick solid curves) and
the $\tilde{M}$-$r_b$ relations (thin dot-dashed curves).
The thin dotted curves indicate the  sequence of the degenerate horizons produced by varying $\tilde{\alpha}$.
We also plot the $\tilde{M}$-$r_h$ relations in general relativity in $n=6$ (thin solid curves) for comparison.
We set $\tilde{\alpha}=0.2$.
The dots with characters ``E" and ``B" imply the degenerate horizon
and the branch point, respectively.
Below the $\tilde{M}$-$r_b$ line, there are no solutions.
The region with a star that is bounded by the $\tilde{M}$-$r_h$ and
$\tilde{M}$-$r_b$ curves is the untrapped region. 
It changes from an untrapped to a trapped region or vice versa by
crossing the $\tilde{M}$-$r_h$ curve.
The $\tilde{M}$-$r$ diagrams of the higher-dimensional solutions 
where $n\geq 6$
have qualitatively similar configurations to these.
}
\label{fig:M-r-6-neutral}
\end{figure}

\begin{figure}[tbp]
\includegraphics[width=.90\linewidth]{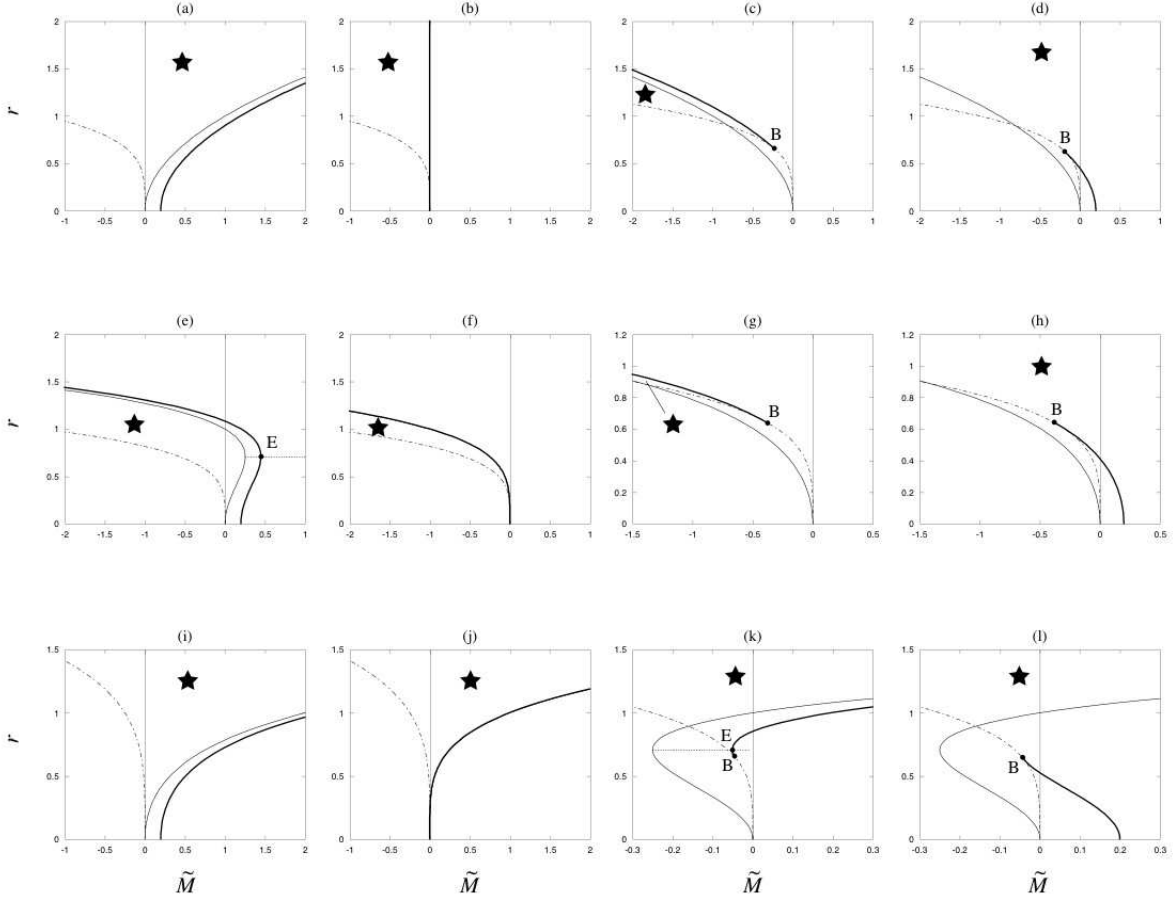}
\caption{
The $\tilde{M}$-$r$ diagrams of the static
solutions in the
five-dimensional Einstein-Gauss-Bonnet-$\Lambda$ system. 
The diagrams in the upper, middle, and lower rows express  the cases where
$1/\ell^2=0$ (zero cosmological constant), 
the $1/\ell^2=-1$ (positive cosmological constant), and
the $1/\ell^2=1$ (negative cosmological constant),
respectively.
The diagrams in the columns on the left-hand side
are the $k=1$ (the minus branch), $0$ (the minus branch), $-1$  (the
minus branch), and  $-1$  (the plus branch) cases, respectively. 
We show
the $\tilde{M}$-$r_h$ relations (thick solid curves)
the $\tilde{M}$-$r_b$ relations (thin dot-dashed curves).
The thin dotted curves indicate the  sequence of the degenerate horizons produced by varying $\tilde{\alpha}$.
We also plot the $\tilde{M}$-$r_h$ relations in general relativity  in $n=5$ (thin solid curves) for comparison.
We set $\tilde{\alpha}=0.2$.
See Fig.~\ref{fig:M-r-6-neutral} for the meanings of
the dots and the stars.
}
\label{fig:M-r-5-neutral}
\end{figure}

\end{widetext}

In the previous section, we discussed properties of the singularity, infinity, and horizons of the solutions in the Einstein-Gauss-Bonnet-$\Lambda$
system. 
In this section we examine the number of horizons and show the spacetime structures of the solutions
by taking into account the above properties.
As an aid to discussion, the $\tilde{M}$-$r$
diagrams are shown in Fig.~\ref{fig:M-r-6-neutral} ($n= 6$)
and Fig.~\ref{fig:M-r-5-neutral} ($n=5$).
The spacetime structures of the solutions are summarized in
Tables~\ref{GB-penrose-6-neutral} ($n=6$) and 
\ref{GB-penrose-5-neutral} ($n=5$).
Since the spacetime structures in 5-dimension are different from
those in the $n\geq 6$ case, we separate the tables.

\vspace{8mm}
\subsection{$\Lambda=0$ case}

\subsubsection{$k=1$  case}

When $k=1$, a solution in the plus branch has no horizon and represents
the spacetime with a globally naked singularity unless $\tilde{M}=0$.
The spacetime with $\tilde{M}=0$  is the $n$-dimensional adS spacetime
whose curvature radius is $\ell_{\rm eff}$. 

In the $n \geq 6$ case,
the $\tilde{M}$-$r$ diagram of the minus branch is shown as in 
Fig.~\ref{fig:M-r-6-neutral}(a).
For $\tilde{M}<0$, the solution has no horizon, and the branch singularity
can be seen from infinity. The solution represents
the spacetime with a globally naked singularity.
For $\tilde{M}=0$, the solution has a regular center, and the spacetime is Minkowski
spacetime.
For $\tilde{M}>0$, the solution has a black hole horizon
and represents the $n$-dimensional Schwarzschild-adS black hole spacetime. 

In the $n =5$ case,
the $\tilde{M}$-$r$ diagram of the minus branch is shown as in 
Fig.~\ref{fig:M-r-5-neutral}(a).
For $\tilde{M}\leq 0$, the diagram is qualitatively the same as that in the
higher-dimensional case.
For $0<\tilde{M}\leq\tilde{\alpha}$ the solution has no horizon and represents
the spacetime with a globally naked singularity. 
For  $\tilde{M}>\tilde{\alpha}$ 
the solution has a black hole horizon
and represents the five-dimensional Schwarzschild-adS black hole spacetime. 
In the $\tilde{M}\to \tilde{\alpha}+$ limit, the black hole horizon is almost degenerate.

\subsubsection{$k=0$  case}

When $k=0$, a solution in the plus branch has no horizon and represents
the spacetime with a globally naked singularity unless $\tilde{M}=0$.

The $\tilde{M}$-$r$ diagrams of the minus branch are shown in 
Figs.~\ref{fig:M-r-6-neutral}(b) and \ref{fig:M-r-5-neutral}(b).
No horizon exists in the minus branch either, and every solution represents
the spacetime with a globally naked singularity.
For $\tilde{M}<0$, there is a branch singularity at $r=r_{b}$. 
For $\tilde{M}>0$,  there is a central singularity at $r=0$. In the whole 
region  the signature of spacetime
changes, i.e., the Killing vector $\partial_t$ is spacelike for $\tilde{M}>0$. We call this type of region a
trapped region.

\subsubsection{$k=-1$  case}

In the $k=-1$, the solution in the plus branch can have a horizon.
First, however,  let us consider a solution in the minus branch.
The $\tilde{M}$-$r$ diagrams  are shown in 
Figs.~\ref{fig:M-r-6-neutral}(c) and \ref{fig:M-r-5-neutral}(c).
The diagrams in $n=5$  and higher dimensional cases are qualitatively the same.
For $\tilde{M}<\tilde{M}_B$, the solution has only a cosmological horizon and represents
the spacetime with a globally naked singularity.
For $\tilde{M}_B\leq\tilde{M}<0$, the solution has  no horizon and represents
the spacetime with a globally naked singularity.
For $\tilde{M}=0$, the solution has  no horizon and represents
the  regular spacetime. However, the signature of spacetime is opposite to that in
the ordinary untrapped region.
For $\tilde{M}>0$, the solution has no horizon and represents
the spacetime with a globally naked singularity. The signature of spacetime is opposite to  that in
the ordinary untrapped region.

In the $n\geq 6$ case,
the $\tilde{M}$-$r$ diagram of the plus branch is shown as in 
Fig.~\ref{fig:M-r-6-neutral}(d).
For $\tilde{M}\leq\tilde{M}_B$, the solution has no horizon and represents
the spacetime with a globally naked singularity.
For $\tilde{M}_B<\tilde{M}<0$, the solution has a black hole horizon. 
It should be noted that the negative mass black hole solution can exist
without a cosmological constant.
For $\tilde{M}=0$, the solution has a black hole horizon and represents
black hole spacetime. This solution is the zero mass black hole.
For $0<\tilde{M}<\tilde{M}_{ex}^{(+)}$, the solution has  two
horizons, which are black hole and inner horizons.  
Hence the spacetime structure
is similar to the RN-adS black hole spacetime, even though the solution is neutral.
In general relativity the negative mass black hole solution always has an
inner horizon, while this solution only has a black hole horizon.
For $\tilde{M}=\tilde{M}_{ex}^{(+)}$, the solution has  a degenerate horizon and
represents the extreme black hole spacetime. 
For $\tilde{M}>\tilde{M}_{ex}^{(+)}$, the solution has  no horizon and represents
the spacetime with a globally naked singularity.

In the $n=5$ case, 
the $\tilde{M}$-$r$ diagram of the plus branch is as shown in 
Fig.~\ref{fig:M-r-5-neutral}(d).
For $\tilde{M}\leq 0$, the diagram is qualitatively the same as that in the
higher-dimensional case.
For $0<\tilde{M}<\tilde{\alpha}$, the solution has a black hole horizon.
In the $\tilde{M}\to \tilde{\alpha}-$ limit, the black hole horizon is almost degenerate.
For $\tilde{M}\geq\tilde{\alpha}$, the solution has no horizon and represents
the spacetime with a globally naked singularity.

\subsection{$\Lambda>0$ case}

\subsubsection{$k=1$  case}

When $k=1$, there is no horizon in the plus branch, and
the spacetime has a globally naked singularity unless $\tilde{M}=0$.
The solution with $\tilde{M}=0$  is the $n$-dimensional dS spacetime. 

In the $n\geq 6$ case,
the $\tilde{M}$-$r$ diagram of the minus branch is shown as in 
Fig.~\ref{fig:M-r-6-neutral}(e).
There is one extreme solution with mass $\tilde{M}_{ex}^{(-)}$ for fixed $\tilde{\alpha}$.
For $\tilde{M}<0$, there is a cosmological horizon and represents
the spacetime with a globally naked singularity. 
For $\tilde{M}=0$, the solution has a regular center and a cosmological horizon.
The spacetime is the $n$-dimensional dS spacetime whose curvature radius is $\sqrt{|\ell_{\rm eff}^2|}$.
For $0<\tilde{M}<\tilde{M}_{ex}^{(-)}$, the solution has a black hole and  a cosmological
horizons. The spacetime is the $n$-dimensional Schwarzschild-dS spacetime.
For $\tilde{M}=\tilde{M}_{ex}^{(-)}$, the solution has  a degenerate horizon and represents  the extreme black hole spacetime.
For $\tilde{M}>\tilde{M}_{ex}^{(-)}$, the solution has  no horizon and represents
the spacetime with a globally naked singularity.

In the $n=5$ case,
the $\tilde{M}$-$r$ diagram of the minus branch is shown in 
Fig.~\ref{fig:M-r-5-neutral}(e).
The diagram is qualitatively the same as that in the
higher dimensional case except for the mass parameter range
$0<\tilde{M}\leq\tilde{\alpha}$.
For $0<\tilde{M}\leq\tilde{\alpha}$, the solution has  a cosmological horizon and represents
the spacetime with a globally naked singularity. 
In the $\tilde{M}\to \tilde{\alpha}+$ limit, the black hole horizon is almost degenerate.

\subsubsection{$k=0$  case}

When $k=0$, there is no horizon in the plus branch, and
the spacetime has a globally naked singularity unless $\tilde{M}=0$.

The $\tilde{M}$-$r$ diagram of the minus branch is shown in 
Figs.~\ref{fig:M-r-6-neutral}(f) and \ref{fig:M-r-5-neutral}(f).
For $\tilde{M}<0$, the solution has a cosmological horizon and represents
the spacetime with a globally naked singularity. 
In the $\tilde{M}\to \tilde{\alpha}-$ limit, the cosmological horizon is almost degenerate.
For $\tilde{M}=0$, the solution has  no horizon and represents the regular spacetime. However,
the spacetime is trapped.
For $\tilde{M}>0$, the solution has  no horizon and represents
the spacetime with a globally naked singularity.

\subsubsection{$k=-1$  case}

The $\tilde{M}$-$r$ diagrams of the minus branches are shown in 
Figs.~\ref{fig:M-r-6-neutral}(g) and \ref{fig:M-r-6-neutral}(h) for $n\geq 6$, and
in Figs.~\ref{fig:M-r-5-neutral}(g) and \ref{fig:M-r-5-neutral}(h)  for $n=5$. 
They are same as those for
$\Lambda=0$ and $k=-1$ qualitatively.  
Hence the classification of the number of the horizons is same.
However, since the
structure of infinity in the minus-branch solution differs according to whether $\Lambda=0$ or $\Lambda>0$,
the spacetime structures shown in 
Tables~\ref{GB-penrose-6-neutral} and \ref{GB-penrose-5-neutral} differ.

\subsection{$\Lambda<0$ case}

\subsubsection{$k=1$  case}

When $k=1$, there is no horizon in the plus branch, and
the solution has a globally naked singularity unless $\tilde{M}=0$.
The spacetime with $\tilde{M}=0$ is the $n$-dimensional adS spacetime. 

In the $n\geq 6$ case,
the $\tilde{M}$-$r$ diagram in the minus branch is shown in as
Fig.~\ref{fig:M-r-6-neutral}(i). 
For $\tilde{M}<0$, the solution has no horizon and represents
the spacetime with a globally naked singularity. 
For $\tilde{M}=0$, the solution has no horizon and represents the
$n$-dimensional adS spacetime. 
For $\tilde{M}>0$, the solution has a black hole horizon and represents
the $n$-dimensional  Schwarzschild-adS black hole. 

In the $n=5$ case,
the $\tilde{M}$-$r$ diagram of the minus branch is shown as in 
Fig.~\ref{fig:M-r-5-neutral}(i). 
The diagram is qualitatively the same as that in the
higher-dimensional case except for the mass parameter range
$0<\tilde{M}\leq\tilde{\alpha}$.
For $0<\tilde{M}\leq\tilde{\alpha}$, the solution has no horizon and represents
the spacetime with a globally naked singularity. 
In the $\tilde{M} \to \tilde{\alpha}+$ limit, the black hole horizon is 
almost degenerate.

\subsubsection{$k=0$  case}

When $k=0$, there is no horizon in the plus branch, and
the spacetime has a globally naked singularity unless $\tilde{M}=0$.

The $\tilde{M}$-$r$ diagram in the minus branch is shown in 
Fig.~\ref{fig:M-r-6-neutral}(j) for $n\geq 6$ case  and in Fig.~\ref{fig:M-r-5-neutral}(j)  for $n= 5$. 
Both diagrams are qualitatively the same.
For $\tilde{M}<0$, the solution has no horizon and represents
the spacetime with a globally naked singularity. 
For $\tilde{M}=0$, the solution has no horizon and represents the regular spacetime. 
For $\tilde{M}>0$, the solution has a black hole horizon. 
In the $\tilde{M} \to 0+$ limit, the black hole horizon is 
almost degenerate.

\subsubsection{$k=-1$  case}

In the $n\geq 6$ case, $r_{ex}$ has two  positive roots by  
Eq.~(\ref{extreme-neutral-1}).
It is easily seen that each extreme solution belongs to the minus and the plus branches, respectively. So we denote their radii as $r_{ex}^{(+)}$ and $r_{ex}^{(-)}$ with $r_{ex}^{(+)}<r_{ex}^{(-)}$.
It is enough to show that $r_{ex}^{(+)}<r_{B}$ and $r_{ex}^{(-)}>r_{B}$.
First let us compare
$r_{ex}^{(+)}$ with $r_{B}$:
\begin{eqnarray}
&&r_B^2-\bigl[r_{ex}^{(+)}\bigr]^2
=\frac{(n-3)\ell^2}{2(n-1)}
\nonumber
\\
&& ~~~~
\times
\biggl[\frac{4(n-1)\tilde{\alpha}}{(n-3)\ell^2}-1
+\sqrt{1-\frac{4\tilde{\alpha}(n-1)(n-5)}{\ell^2(n-3)^2}}
\biggr].
\nonumber
\\
\end{eqnarray}  
If the first term in the second line is larger than or equal to $1$, the rhs
of the equation is always positive. Hence we find $r_{ex}^{(+)}<r_{B}$.
If not, the sign of the rhs is determined by the sum of
the first two terms and the third term. Calculating
\begin{eqnarray}
&&\biggl[1-\frac{4\tilde{\alpha}(n-1)(n-5)}{\ell^2(n-3)^2}\biggr]
-\biggl[\frac{4(n-1)\tilde{\alpha}}{(n-3)\ell^2}-1\biggr]^2
\nonumber
\\
&& ~~~~
=\frac{4\tilde{\alpha}(n-1)^2}{\ell^2(n-3)^2}
\biggl(1-\frac{4\tilde{\alpha}}{\ell^2}\biggr)>0,
\end{eqnarray}  
we find again the $r_{ex}^{(+)}$ is less than $r_B$.
This implies that
the extreme solution with $r_{ex}^{(+)}$ belongs to the plus
branch. 

Next, let us show $r_{ex}^{(-)}>r_B$. 
Equation~(\ref{extreme-neutral-1}) implies that $r_{ex}^{(-)}$ is the
function of $\tilde{\alpha}$. 
$r_{ex}^{(-)}$ monotonically increases as $\tilde{\alpha}$ increases. 
For  $4\tilde{\alpha}/\ell^2=1$, the radius of the degenerate
horizon is $r_{ex}^{(-)}=r_B$.  
This implies that $r_{ex}^{(-)}>r_B$ for $4\tilde{\alpha}/\ell^2<1$.

In the $n=5$ case, $r_{ex}$ has only one positive root 
by Eq.~(\ref{extreme-neutral-1}).
It is also shown in the similar way that the extreme solution belongs to the
minus branch.

Let us consider a solution of the minus branch.
The $\tilde{M}$-$r$ diagrams  are shown in 
Figs.~\ref{fig:M-r-6-neutral}(k) for $n\geq 6$ case and \ref{fig:M-r-5-neutral}(k)  for  $n=5$ case.
The diagrams of $n=5$  and higher-dimensional cases are qualitatively the same.
For $\tilde{M}<\tilde{M}_{ex}^{(-)}$, the solution has  no horizon and represents
the spacetime with a globally naked singularity.
For $\tilde{M}=\tilde{M}_{ex}^{(-)}$, the solution has a degenerate horizon and
represents the  extreme black hole spacetime.
For $\tilde{M}_{ex}^{(-)}<\tilde{M}<\tilde{M}_B$, 
the solution has a black hole and an inner horizons. 
For $\tilde{M}_B\leq\tilde{M}<0$, 
there is a black hole horizon. This negative mass black hole solution has no
inner horizon.
For $\tilde{M}=0$, the solution has a black hole horizon and a regular center.
For $\tilde{M}>0$, the solution has a black hole horizon.

In the $n\geq 6$ case,
the $\tilde{M}$-$r$ diagram of the plus branch is shown as in 
Fig.~\ref{fig:M-r-6-neutral}(l).
For $\tilde{M}\leq\tilde{M}_B$, the solution has no horizon and represents
the spacetime with a globally naked singularity.
For $\tilde{M}_B<\tilde{M}<0$, the solution has a black hole horizon.
For $\tilde{M}=0$, the solution has a black hole horizon and a regular center.
For $0<\tilde{M}<\tilde{M}_{ex}^{(+)}$, the solution has a black hole and an inner
horizons.
For $\tilde{M}=\tilde{M}_{ex}^{(+)}$, the solution has a degenerate horizon and represents the  extreme black hole spacetime.
For $\tilde{M}>\tilde{M}_{ex}^{(+)}$, the solution has no horizon and represents
the spacetime with a globally naked singularity.

In the $n=5$ case
the $\tilde{M}$-$r$ diagram in the plus branch is shown as in 
Fig.~\ref{fig:M-r-5-neutral}(l).
The diagram is qualitatively the same as that in the
higher-dimensional case except for the positive mass parameter range.
For $0<\tilde{M}<\tilde{\alpha}$, the solution has a black hole horizon. 
For $\tilde{M}\geq\tilde{\alpha}$, the solution has no horizon and represents
the spacetime with a globally naked singularity. 
In the $\tilde{M} \to \tilde{\alpha}-$ limit, the black hole horizon is 
almost degenerate.

\section{Einstein-Gauss-Bonnet-$\Lambda$ system:
$4\tilde{\alpha}/\ell^2=1$ case}
\label{EGBL2}

The system in which $4\tilde{\alpha}/\ell^2=1$ is a special case,
where the plus and the minus branches coincide in the sourceless case $M=0$. 
Since we have assumed that $\alpha$ is positive,
here only a negative $\Lambda$ is allowed. 
The metric function $f$ becomes
\begin{equation}
f=k+\frac{r^2}{2\tilde{\alpha}}
\Biggl(1\mp\sqrt{\frac{4\tilde{\alpha}\tilde{M}}{r^{n-1}}} \Biggr).
\end{equation}
In the large radius the spacetime approaches the adS spacetime  for $r\to \infty$ with the
effective curvature radius $\ell_{\rm eff}:=\sqrt{2\tilde{\alpha}}$ for $k=1$.


The mass $\tilde{M}$ cannot
be negative since the metric function $f$ takes a complex value. 
The boundary $\tilde{M}=0$ 
corresponds to the condition of the branch singularity for 
$4\tilde{\alpha}/\ell^2 \ne 1$. In this sense it can be said that there is a branch singularity at $\tilde{M}=0-$ in the $\tilde{M}$-$r$ diagram. For $\tilde{M}=0$, the solution is
the adS spacetime when $k=1$. For positive $\tilde{M}$, there is a curvature singularity at $r=0$, where the
Kretschmann invariant behaves as
\begin{equation}
{\cal I}=O(r^{-(n-1)}).
\end{equation}
When $n\geq 6$,
the singularity of the minus- (plus-) branch solution is 
spacelike (Fig.~\ref{penrose-center}(iv)) 
(timelike (Fig.~\ref{penrose-center}(i))).
When $n=5$, the structure of the singularity is  as complicated
as that where $4\tilde{\alpha}/\ell^2\ne 1$.
In the minus branch, the singularity is spacelike  when $k=0$ and
$-1$. When $k=1$, the singularity is spacelike  for 
$\tilde{M}> \tilde{\alpha}$, null (Fig.~\ref{penrose-center}(ii)) for $\tilde{M}= \tilde{\alpha}$,
and timelike  for $\tilde{M}< \tilde{\alpha}$.
In the plus branch, the singularity is timelike  when $k=0$ and $1$.
When $k=-1$, the singularity is spacelike  for 
$0<\tilde{M}< \tilde{\alpha}$, null  for $\tilde{M}= \tilde{\alpha}$,
and timelike  for $\tilde{M}> \tilde{\alpha}$.


Since the radius of a horizon  is restricted to
$r_h^2<-2\tilde{\alpha}k$ ($r_h^2>-2\tilde{\alpha}k$) for the plus (minus) branch,
there are no horizons
for solutions with $k=0,~1$ in the plus branch.
The $\tilde{M}$-$r_h$ relation Eq.~(\ref{M-horizon}) becomes simple in form:
\begin{equation}
\tilde{M}=
\frac{r_h^{n-1}}{4\tilde{\alpha}}
\biggl(1+\frac{2k\tilde{\alpha}}{r_h^2}\biggr)^2.
\end{equation}
The $\tilde{M}$-$r_h$ curve behaves $\tilde{M}\to 0$ as $r_h\to 0$ for $n\geq 6$,
and  $\tilde{M}\to |k|\tilde{\alpha}$ for $n=5$.

The relation between the vertical points of the $\tilde{M}$-$r_h$ curve and extreme solutions is the same as that for $4\tilde{\alpha}/\ell^2\ne 1$.
From the condition 
$f=df/dr=0$, the degenerate horizon appears as
\begin{equation}
\label{spe-ex-r}
r_{ex}=\sqrt{-\frac{2(n-5)\tilde{\alpha}k}{n-1}}
\end{equation}
for  $k=-1$ where $n\geq 6$. 
Because $r_{ex}<\sqrt{2\tilde{\alpha}}$, the extreme solution
appears only in the plus branch. 
The mass of the extreme solution is 
\begin{equation}
\label{spe-ex-M}
\tilde{M}_{ex}^{(+)}=\frac{4r_{ex}^{n-1}}{(n-5)^2\tilde{\alpha}}.
\end{equation}
The almost degenerate horizon appears in the
$r_h \to 0$ limit for $k=0$ where $n\geq 6$ and for any $k$ where $n=5$.

\begin{figure}
\includegraphics[width=.99\linewidth]{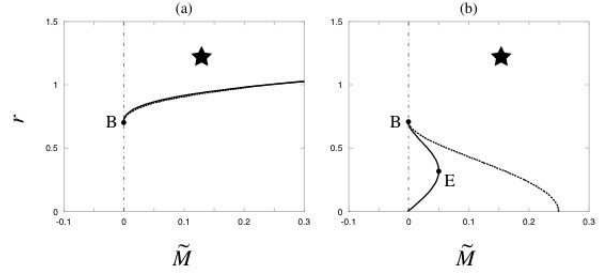}
\caption{
The $\tilde{M}$-$r$ diagrams of the static
solutions in the
Einstein-Gauss-Bonnet-$\Lambda$ system
where $4\tilde{\alpha}/\ell^2=1$. 
The diagrams  are 
(a) the minus branch and (b) the plus branch for $k=-1$.
The solid  and dashed curves are for $n=6$ and $n=5$, respectively.
In (a), both curves overlap each other.
See Figs.~\ref{fig:E_m-rh} and \ref{fig:M-r-6-neutral}  for the
meanings of the dots and the stars.
}
\label{fig:M-r-spe}
\end{figure}


When $k=1,~0$, the $\tilde{M}$-$r$ diagrams
are the same as those for $4\tilde{\alpha}/\ell^2\ne 1$
qualitatively (Figs.~\ref{fig:M-r-6-neutral}(i), (j) and
\ref{fig:M-r-5-neutral}(i), (j)) except that negative mass
solutions are forbidden. 
For $\tilde{M}=0$, the solution has no horizon, and the spacetime is regular everywhere.
The solution where $k=1$ is the adS solution whose curvature radius is $\ell_{\rm  eff} = \sqrt{2\tilde{\alpha}}$.
For $\tilde{M}>0$, the solution has a black hole horizon. 
In the $k=0$ and/or $n=5$ case, 
the infinitesimally small  black hole has an almost degenerate horizon.

When $k=-1$, the $\tilde{M}$-$r_h$ diagram in the minus branch is shown as in
Fig.~\ref{fig:M-r-spe}(a). 
The negative mass range  $\tilde{M}<0$ is forbidden.
For $\tilde{M}=0$, the solution has a black hole horizon, and the spacetime is regular everywhere.
For $\tilde{M}>0$, the solution has a black hole horizon. 
Even in the infinitesimally small mass limit, the horizon radius remains finite:
$r_h\approx \sqrt{2\tilde{\alpha}}$.

The $\tilde{M}$-$r_h$ diagram in the plus branch for $k=-1$ is shown in
Fig.~\ref{fig:M-r-spe}(b). 
In the $n\geq 6$ case, there is one extreme solution where mass and radius are
given by
Eqs.~(\ref{spe-ex-M}) and (\ref{spe-ex-r}), respectively.
For  $\tilde{M}=0$, the solution has a black hole horizon and  represents the regular spacetime. 
For $0<\tilde{M}<M_{ex}^{(+)}$, the solution has an inner and a black hole horizons and represents the  black hole spacetime. 
For $\tilde{M}=M_{ex}^{(+)}$, the solution has a degenerate horizon and represents the extreme black hole spacetime.
For $\tilde{M}>M_{ex}^{(+)}$, the solution has no horizon and represents the spacetime with a  globally naked singularity.
In the $n=5$ case, there is no extreme solution.
For  $\tilde{M}=0$, the solution has a black hole horizon, and  the spacetime is regular everywhere.
For $0<\tilde{M}<\sqrt{\alpha}$, the solution has  a black hole horizon, represents the  black hole spacetime. In the $\tilde{M}\to\sqrt{\alpha}-$ limit,
the horizon is almost degenerate.
For $\tilde{M}\geq\sqrt{\alpha}$, the solution has no horizon and represents the spacetime with a globally naked singularity.
The spacetime structures of these solutions are summarized in
Table~\ref{GB-penrose-spe-neutral}.


\section{Conclusions and Discussion}
\label{Conclusion}

We have studied spacetime structures of the static solutions in
the $n$-dimensional Einstein-Gauss-Bonnet-$\Lambda$ system systematically. 
We assume that the Gauss-Bonnet coefficient $\alpha$ is non-negative. This assumption is consistent with the low energy effective action derived from superstring/M-theory.
The solutions we considered in this paper have the $(n-2)$-dimensional Euclidean sub-manifold, which is the Einstein, for example, sphere, flat, or hyperboloid.
When the curvature of the sub-manifold is $k=0,~-1$, black holes are called topological black holes. 
We assume $4{\tilde \alpha}/\ell^2\leq 1$ in order for the sourceless solution  ($M=0$) to be defined.
There is no solution with a negative mass parameter when $4{\tilde \alpha}/\ell^2=1$. 
The structures of the center, horizons, asymptotic regions, and the singular point depend on the parameters $\alpha$, $\ell^2$, $k$, $M$, and branches complicatedly so that a variety of global structures for the solution are found. 
In our analysis, an $\tilde{M}$-$r$ diagram is used on which the $\tilde{M}$-$r_h$,
$\tilde{M}$-$r_{ex}$ and $\tilde{M}$-$r_b$ curves are drawn. It makes our consideration clear and enables easy understanding through visual effects. The solutions are classified into regular, black hole and globally naked solutions.

In Gauss-Bonnet gravity, the general solutions, which were first obtained by Boulware and Deser for the vanishing cosmological constant and for $k=1$, are classified into the plus and the minus branches. In the $\tilde{\alpha}\to 0$ limit, 
the solution in the minus branch recovers the one in general relativity, while there is no solution in the plus branch.
In general relativity, the behavior of $r \to \infty$ is dominated by the cosmological constant, and the asymptotic structure  of a solution is generally determined by that term. In Gauss-Bonnet gravity, the curvature radius  $\ell^2=-(n-1)(n-2)/2\Lambda$ in general relativity is replaced by the effective curvature radius 
$\ell_{\rm eff}^2$. In the minus branch the signs of $\ell_{\rm eff}^2$ and $\ell^2$ are same, and the ordinary correspondence between the signs of the cosmological constant and the asymptotic structures is obtained. In the plus branch, however, the signs
of $\ell_{\rm eff}^2$ are always positive independently of the cosmological constant,
and all the solutions have the same  asymptotic structure as that in general relativity with a negative cosmological constant.

For the positive mass parameter, the singularity exists at the center. There the Kretschmann invariant behaves as $O(r^{-(n-1)})$. 
For the negative mass parameter, the new type of singularity called  the branch singularity appears at non-zero finite 
radius $r=r_b>0$. There the Kretschmann invariant behaves as $O((r-r_b)^{-3})$. In any case, the divergent behavior in Gauss-Bonnet gravity is milder than the central singularity in general relativity $O(r^{-2(n-1)})$.

Among the solutions in our analysis, the black hole solutions are the most important.
There are three types of horizon: inner,  black hole, and  cosmological horizons.
In the $k=1,~0$ cases the plus branch solutions do not have any horizon.
In the $k=-1$ case, the radius of the horizon is restricted as $r_h<\sqrt{2\tilde{\alpha}}$ ($r_h>\sqrt{2\tilde{\alpha}}$) in the plus (minus)
branch. The black hole solution in the plus branch for $k=-1$ has interesting properties.
In general relativity, the black hole solution with zero or negative mass appears
only in the negative cosmological constant system, while such a solution exists in the
plus branch
even with the zero or positive cosmological constant in Gauss-Bonnet gravity.
There are also the extreme black hole solutions with positive mass in spite of the
lack of an electromagnetic charge. 

For $k= 1$, the horizon radius of the black hole solutions in Gauss-Bonnet gravity
is smaller than that in general relativity. Furthermore,
in the $n=5$ case, the infinitesimally small black hole solution has a finite mass 
$\tilde{M}=\tilde{\alpha}$ below which a black hole solution does not exist.
This behavior would affect the black hole formation by a linear collider. 
Recent investigation of the higher-dimensional models represented by a braneworld model predicts such black hole formation if the fundamental energy scale is
small enough. A rough estimate of these analyses is based on the Schwarzschild radius of the colliding particles. If their Schwarzschild radii calculated by the 
higher-dimensional model are comparable to the size of the colliding particles, a black hole will
be formed. By taking into account the Gauss-Bonnet corrections, however, the 
Schwarzschild radius becomes small. This implies that the black hole formation occurs less frequently or not at all \cite{Rychkov}.

\begin{figure}
\includegraphics[width=1.00\linewidth]{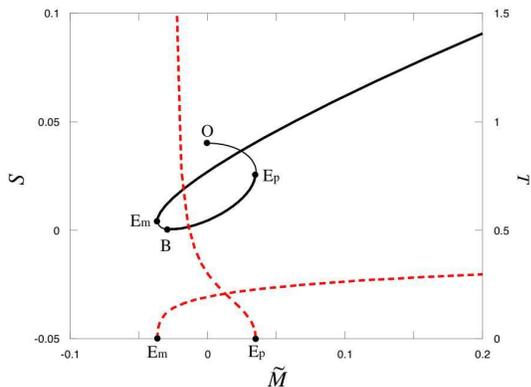}
\caption{
The $\tilde{M}$-$S$ (solid curve) and $\tilde{M}$-$T$  (dashed curves) diagram of the black hole
solutions with $1/\ell^2=1$ and $k=-1$ in the six-dimensional
Einstein-Gauss-Bonnet-$\Lambda$ system.
We set $\tilde{\alpha}=0.2$.
The thick solid curve ending at point ${\rm E_m}$ shows the entropy of the black hole
in the minus branch, and the thick solid curve connecting points B and  
${\rm E_p}$ shows it in the plus branch. Points ${\rm E_m}$ and ${\rm E_p}$ indicate the 
extreme solutions. The thin solid curves between ${\rm E_m}$ and B, and ${\rm E_p}$ and O are the ``entropy" of the inner horizon calculated by the analogy of the black hole horizon case. 
Although the radius of the black hole horizon decreases as the
mass  becomes large in the plus branch,
its entropy increases. This behavior is consistent with the second law of the black hole thermodynamics. 
The dashed curve with ${\rm E_m}$ (${\rm E_p}$) at the end shows  the temperature
of the black hole in the minus (plus) branch. At  ${\rm E_m}$ and 
${\rm E_p}$, the temperature becomes zero, while at point B, it diverges.
}
\label{Fig-M-T}
\end{figure}

Although we focus on the spacetime structure of the solutions in this paper, the thermodynamical properties of the black hole solution are important, in particular in the discussions of the evolution  through Hawking radiation and final state of the black hole spacetime.
The temperature of a black hole in Gauss-Bonnet gravity is obtained by the periodicity of the Euclidean time on the
horizon by
\begin{eqnarray}
T=\frac{(n-1)r_h^4/\ell^2+(n-3)k r_h^2+(n-5)\tilde{\alpha}k^2}
{4\pi r_h(r_h^2+2\tilde{\alpha}k)}.
\;\;\;
\end{eqnarray}  
In general relativity, the horizon radius of the 
black hole is related to the entropy of the black hole $S$
as $S=\pi r_h^2$. 
Since the temperature is $T=dM/dS$
from the first law of the black hole thermodynamics, 
a black hole with its horizon at the vertical point of the $\tilde{M}$-$r_h$ curve
where the horizon is degenerate has zero temperature.
A solution with the almost degenerate horizon has infinitesimally small temperature.
Although the expression of the entropy in Gauss-Bonnet gravity is different from  that in general relativity, it is easy to show that the solution, at which the $\tilde{M}$-$r_h$ curve becomes vertical on the diagram, where its horizon degenerates, has zero temperature.
In Gauss-Bonnet gravity, entropy is not obtained by
a quarter of the area of a black hole horizon.
The second law of black hole  thermodynamics would hold, while
the area theorem would not.
The relevant entropy of the system where the curvature terms exist in the action 
besides the Einstein-Hilbert action was proposed by Iyer and Wald, who regard it as the Noether charge associated
with the diffeomorphism invariance \cite{Iyer}. It is calculated as 
\begin{eqnarray}
S
=\frac{r_h^{n-2}\; \Sigma^k_{n-2}}{4G_n}
\biggl[1+\frac{2(n-2)\tilde{\alpha}k}{(n-4)r_h^2}\biggr]-S_{\rm min},
\end{eqnarray}  
in our model. $S_{\rm min}$ is added to make the entropy non-negative \cite{Clunan}.

In the $n=5$ case, a black hole solution with $k=1$ has the minimum mass $\tilde{\alpha}$. As the evaporation proceeds, the black hole loses its mass and
its horizon radius shrinks to zero, 
and then  the black hole seems to evolve to a naked singularity with the mass $\tilde{M}=\tilde{\alpha}$.
However, since the temperature of the horizon becomes zero in this limit, we cannot be sure 
whether the black hole evaporates thoroughly and evolves to a singularity or not. Similar situations occur in the $\Lambda =-1$ and $k=0$ case. We need detailed analysis of the evaporation process to clarify this issue.

Where $\Lambda =-1$ and $k=-1$, the case is more interesting. There are black hole solutions both in the plus and the minus branches. In the minus branch, the black hole loses its mass, and the horizon radius
becomes small through evaporation. As the solution approaches the extreme solution, evaporation becomes weak, and the black hole may evolve to the extreme solution in an infinite time.
On the other hand, for the black hole solution in the plus branch, as the black hole loses
its mass, the black hole horizon becomes large.
Since the temperature of the black hole solution 
around the branch points B in Figs.~\ref{fig:M-r-6-neutral} and \ref{fig:M-r-5-neutral} is non-zero, the black hole will evolve to a naked singularity that is not point-like but , rather, locates at $r=r_b$.
Conversely, the horizon area
becomes small when matter is put into the black hole. This means that the 
area theorem in general relativity does not extend to Gauss-Bonnet gravity.
However, it is confirmed that entropy increases in such  a classical process, as is shown in Fig.~\ref{Fig-M-T}.
In general relativity there is  an extreme limit of the Reissner-Nordstr\"om black hole. In this case, if we throw matter into the black hole, the ratio of the charge to the mass  of the solution decreases, and it is impossible to make the solution extreme.
If we continue to throw matter into our black hole, the solution will evolve to the
extreme solution. As a result the mass exceeds that of the extreme solution, and
the black hole becomes a naked singularity. This would be a violation of the third law of black hole thermodynamics.

As an important application of the solutions studied in this paper, it is known that the static black hole or the regular solutions with the metric of Eq.~(\ref{metric}) can be used as a bulk spacetime in the braneworld cosmology. 
In the braneworld scenario, our universe is expressed by a shell in the 
higher-dimensional bulk spacetime. 
The radius of the shell in the bulk spacetime corresponds to the scale factor of our universe, so that the motion of the shell gives the evolution of the universe. 
In the Einstein-Maxwell-$\Lambda$ system, an infalling shell in the Reissner-Nordstr\"om bulk spacetime bounces at some finite radius~\cite{RN-brane}. 
This motion gives the bouncing universe without the big-bang singularity. 
However, this bounce occurs only inside of the inner horizon.  
As is well known, the inner horizon is unstable, so that the bouncing universe is hardly to be realized~\cite{mass-inf,kink}. 
On the other hand, although there are several investigations of the braneworld in Gauss-Bonnet gravity \cite{GBbrane}, much is not known.
The bounce may occur outside of the inner horizon in Gauss-Bonnet gravity, so that a non-singular cosmology could be possible.
These are under investigation~\cite{Torii-brane}.

\section*{Acknowledgements}

We would like to thank 
Kei-ichi Maeda 
and 
Umpei Miyamoto 
for their discussion. 
This work was partially supported by a Grant for The 21st Century COE Program (Holistic Research and Education Center for Physics Self-Organization Systems) at Waseda University.


\vspace{10mm}
\begin{figure}[h]
\includegraphics[width=.90\linewidth]{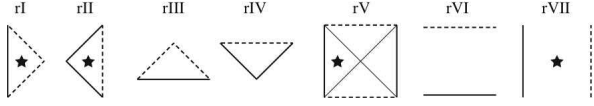}
\caption{
The conformal diagrams of regular solutions.
The region with a star is the untrapped region where the Killing
vector $\partial_t$ is timelike. It changes from the untrapped to the trapped region
or vise versa by crossing the horizon. 
In the following figures, we use the same mark.
rI, rV, and rVII are Minkowski, dS
and adS spacetimes, respectively.
There are time-reversed diagrams for rIII, rIV, and rVI.
}
\label{penrose_regular}
\end{figure}

\begin{figure}[h]
\includegraphics[width=.99\linewidth]{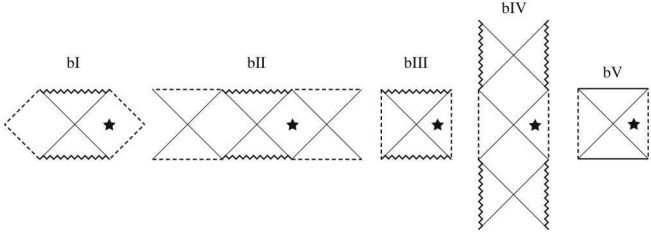}
\caption{
The conformal diagrams of the black hole solutions.
bI, bII, bIII, and bIV have the same structure as those of Schwarzschild, 
Schwarzschild-dS, Schwarzschild-adS,
and RN-adS spacetimes, respectively.
}
\label{penrose_BH}
\end{figure}

\begin{figure}[h]
\includegraphics[width=.85\linewidth]{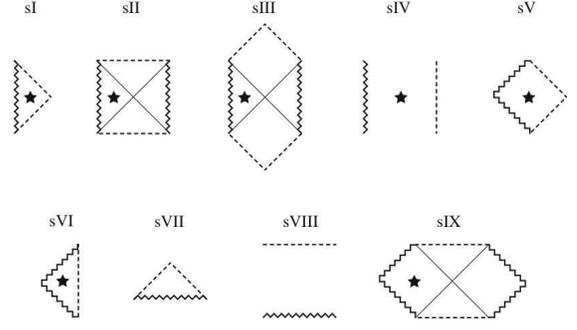}
\caption{
The conformal diagrams of the solutions with a globally naked singularity.
There are time-reversed diagrams for sVII and sVIII.
}
\label{penrose_singular}
\end{figure}

\begin{figure}[t]
\includegraphics[width=.48\linewidth]{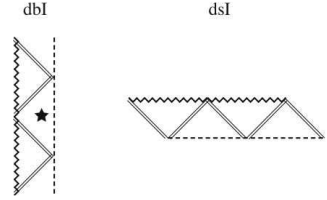}
\caption{
The conformal diagrams of (dbI) the  black hole solution
with degenerate horizons and (dsI)  the  solutions  with a globally naked singularity
and degenerate horizons.
The degenerate horizon is drawn with a double line. By passing across the double
line, the trapped (untrapped) region does not change to a untrapped (trapped) 
one.
dbI has the same structure as that of the extreme RN-adS spacetimes.
There is a time-reversed diagram for dsI.
}
\label{penrose_BH_degenerate}
\end{figure}

\newpage

\begin{widetext}

\begin{table}
\renewcommand{\arraystretch}{1.2}
\caption{
The classification of the spacetime structures of the static
solutions in the Einstein-$\Lambda$ system.
The numbers in the column ``type" imply the types of the conformal
diagrams in 
Figs.~\ref{penrose_regular}--\ref{penrose_BH_degenerate}.
}
\label{E-penrose}
\vspace{4mm}
\begin{tabular}{c|c|c|c|c|c|c}
\hline
\hline
\lw{~$k$~} &
\multicolumn{2}{c|}{$\Lambda=0$} & 
\multicolumn{2}{c|}{$\Lambda>0$} &
\multicolumn{2}{c}{$\Lambda<0$}
\\
\cline{2-7}
&  ~~~~~~~~~~$\tilde{M}$~~~~~~~~~~ & ~type~ &
~~~~~~~~~~$\tilde{M}$~~~~~~~~~~ & ~type~ &
~~~~~~~~~~$\tilde{M}$~~~~~~~~~~ & ~type~ 
\\
\hline \hline 
 1 & $\tilde{M}<0$ & sI  &
$\tilde{M}<0$ & sII  &
$\tilde{M}<0$ & sIV
\\
\cline{2-7}
   & $\tilde{M}=0$ & rI &
$\tilde{M}=0$ & rV  &
$\tilde{M}=0$ & rVII
\\
\cline{2-7}
   & $\tilde{M}>0$ & bI &
$0<\tilde{M}<\tilde{M}_{ex}$ & bII  &
$\tilde{M}>0$ & bIII
\\
\cline{2-7}
    &  &    &
$\tilde{M}=\tilde{M}_{ex}$ & dsI  &
    &      
\\
\cline{4-5}
   &  &   &
$\tilde{M}>\tilde{M}_{ex}$ & sVIII  &
    &      
\\
\noalign{\hrule height 1.0pt}
 0  & $\tilde{M}<0$ &  sI &
$\tilde{M}<0$ &  sII  &
$\tilde{M}<0$ & sIV
\\
\cline{2-7}
   & $\tilde{M}=0$ & --- &
$\tilde{M}=0$ & rIV  &
$\tilde{M}=0$ & rII
\\
\cline{2-7}
   & $\tilde{M}>0$ & sVII &
$\tilde{M}>0$ & sVIII  &
$\tilde{M}>0$ & bIII
\\
\noalign{\hrule height 1.0pt}
 -1  & $\tilde{M}<0$ &  sIII &
$\tilde{M}<0$ &  sII &
$\tilde{M}<\tilde{M}_{ex}$ &  sIV
\\
\cline{2-7}
   & $\tilde{M}=0$ & rIII &
$\tilde{M}=0$ & rVI &
$\tilde{M}=\tilde{M}_{ex}$ & dbI
\\
\cline{2-7}
   & $\tilde{M}>0$ & sVII &
$\tilde{M}>0$ & sVIII &
$\tilde{M}_{ex}<\tilde{M}<0$ & bIV
\\
\cline{2-7}
 & & &
 & &
 $\tilde{M}=0$ & bV
\\
\cline{6-7}
 & & &
 & &
 $\tilde{M}>0$ & bIII
\\
\hline
\hline
\end{tabular}
\end{table}



\begin{table}
\renewcommand{\arraystretch}{1.2}
\caption{ 
The classification of the spacetime structures of the static solutions in the $n\geq 6$ Einstein-Gauss-Bonnet-$\Lambda$
system.  The
numbers in the column ``type" imply the types of the conformal diagrams in
Figs.~\ref{penrose_regular}-\ref{penrose_BH_degenerate}.
}
\label{GB-penrose-6-neutral}
\vspace{4mm}
\begin{tabular}{c|c|c|c|c|c|c|c|c}
\hline
\hline
\lw{~$k$~} &
\multicolumn{2}{c|}{$\Lambda=0$, ($-$ branch)} & 
\multicolumn{2}{c|}{$\Lambda>0$, ($-$ branch)} &
\multicolumn{2}{c|}{$\Lambda<0$, ($-$ branch)} &
\multicolumn{2}{c}{$+$ branch} 
\\
\cline{2-9}
&  ~~~~~~~~~~$\tilde{M}$~~~~~~~~~~ & ~type~ 
&   ~~~~~~~~~~$\tilde{M}$~~~~~~~~~~ & ~type~ 
&   ~~~~~~~~~~$\tilde{M}$~~~~~~~~~~ & ~type~ 
& ~~~~~~~~~~$\tilde{M}$~~~~~~~~~~ & ~type~ 
\\
\hline \hline 
1   & $\tilde{M}<0$ & sI 
& $\tilde{M}<0$ & sII 
& $\tilde{M}<0$ & sIV 
& $\tilde{M}\ne 0$ & sIV 
\\
\cline{2-9}
& $\tilde{M}=0$ & rI
& $\tilde{M}=0$ & rV 
& $\tilde{M}=0$ & rVII 
& $\tilde{M}=0$ & rVII 
\\
\cline{2-9}
& $\tilde{M}>0$ & bI
& $0<\tilde{M}<\tilde{M}_{ex}^{(-)}$ & bII 
& $\tilde{M}>0$ & bIII 
\\
\cline{2-7}
&  & 
& $\tilde{M}=\tilde{M}_{ex}^{(-)}$ & dsI 
&  &  
\\
\cline{4-5}
&  & 
& $\tilde{M}>\tilde{M}_{ex}^{(-)}$ & sVIII 
&  &
\\
\noalign{\hrule height 1.0pt}
0  & $\tilde{M}<0$ & sI
& $\tilde{M}<0$ & sII 
& $\tilde{M}<0$ & sIV 
& $\tilde{M}\ne 0$ & sIV
\\
\cline{2-9}
& $\tilde{M}=0$ & ---
& $\tilde{M}=0$ & rIV 
& $\tilde{M}=0$ & rII 
& $\tilde{M}=0$ & rII 
\\
\cline{2-9}
& $\tilde{M}>0$ & sVII
& $\tilde{M}>0$ & sVIII
& $\tilde{M}>0$ & bIII 
\\
\noalign{\hrule height 1.0pt}
-1  & $\tilde{M}<\tilde{M}_{B}$ & sIII
& $\tilde{M}<\tilde{M}_{B}$ & sII 
& $\tilde{M}<\tilde{M}_{ex}^{(-)}$ & sIV 
& $\tilde{M}<\tilde{M}_{B}$ & sIV
\\
\cline{2-9}
& $\tilde{M}=\tilde{M}_{B}$ & sVII
& $\tilde{M}=\tilde{M}_{B}$ & sVIII
&  $\tilde{M}=\tilde{M}_{ex}^{(-)}$ & dbI 
& $\tilde{M}=\tilde{M}_{B}$ & sIV
\\
\cline{2-9}
& $\tilde{M}>\tilde{M}_{B}$ $(\tilde{M}\ne 0)$ & sVII
& $\tilde{M}>\tilde{M}_{B}$ $(\tilde{M}\ne 0)$ & sVIII 
&  $\tilde{M}_{ex}^{(-)}<\tilde{M}<\tilde{M}_{B}$ & bIV 
& $\tilde{M}_{B}<\tilde{M}<0$ & bIII
\\
\cline{2-9}
& $\tilde{M}= 0$ & rIII
& $\tilde{M}= 0$& rVI 
&  $\tilde{M}=\tilde{M}_{B}$ & bIII 
& $\tilde{M}=0$ & bV
\\
\cline{2-9}
&  &
&  &
&  $\tilde{M}>\tilde{M}_{B}$ $(\tilde{M}\ne 0)$ & bIII
& $0<\tilde{M}<\tilde{M}_{ex}^{(+)}$ & bIV
\\
\cline{6-9}
&  &
&  &
& $\tilde{M}=0$  & bV
& $\tilde{M}=\tilde{M}_{ex}^{(+)}$ & dbI
\\
\cline{6-9}
&  &
&  &
&  &
& $\tilde{M}>\tilde{M}_{ex}^{(+)}$ & sIV
\\
\hline
\hline
\end{tabular}
\vspace{18pt}
\end{table}



\begin{table}
\renewcommand{\arraystretch}{1.2}
\caption{ 
The classification of the spacetime structures of the static solutions in the five-dimensional Einstein-Gauss-Bonnet-$\Lambda$
system.  The
numbers in the column ``type" imply the types of the conformal diagrams
in Figs.~\ref{penrose_regular}-\ref{penrose_BH_degenerate}. }
\label{GB-penrose-5-neutral}
\vspace{4mm}
\begin{tabular}{c|c|c|c|c|c|c|c|c}
\hline
\hline
\lw{~$k$~} 
&\multicolumn{2}{c|}{$\Lambda=0$ ($-$ branch)} 
&\multicolumn{2}{c|}{$\Lambda>0$ ($-$ branch)} 
&\multicolumn{2}{c|}{$\Lambda<0$ ($-$ branch)}
&\multicolumn{2}{c}{$+$ branch}
\\
\cline{2-9}
&  ~~~~~~~~~~$\tilde{M}$~~~~~~~~~~ & ~type~ 
&  ~~~~~~~~~~$\tilde{M}$~~~~~~~~~~ & ~type~ 
&  ~~~~~~~~~~$\tilde{M}$~~~~~~~~~~ & ~type~ 
&  ~~~~~~~~~~$\tilde{M}$~~~~~~~~~~ & ~type~ 
\\
\hline \hline 
1 & $\tilde{M}<\tilde{\alpha}$ $(\tilde{M}\ne 0)$ & sI 
& $\tilde{M}<\tilde{\alpha}$ $(\tilde{M}\ne 0)$ & sII 
& $\tilde{M}<\tilde{\alpha}$ $(\tilde{M}\ne 0)$ & sIV
& $\tilde{M}\ne 0$ & sIV
\\
\cline{2-9}
& $\tilde{M}=0$ & rI
& $\tilde{M}=0$ & rV 
& $\tilde{M}=0$ & rVII 
& $\tilde{M}=0$ & rVII 
\\
\cline{2-9}
& $\tilde{M}=\tilde{\alpha}$ & sV
& $\tilde{M}=\tilde{\alpha}$ & sIX 
& $\tilde{M}=\tilde{\alpha}$ & sVI 
&  & 
\\
\cline{2-9}
& $\tilde{M}>\tilde{\alpha}$ & bI
& $\tilde{\alpha}<\tilde{M}<\tilde{M}_{ex}^{(-)}$ & bII 
& $\tilde{M}>\tilde{\alpha}$ & bIII 
\\
\cline{2-7}
&  & 
& $\tilde{M}=\tilde{M}_{ex}^{(-)}$ & dsI 
&  & 
\\
\cline{4-5}
&  & 
& $\tilde{M}>\tilde{M}_{ex}^{(-)}$ & sVIII 
&  & 
\\
\noalign{\hrule height 1.0pt}
0 &  $\tilde{M}<0$ & sI
& $\tilde{M}<0$ & sII 
& $\tilde{M}<0$ & sIV 
& $\tilde{M}\ne 0$ & sIV
\\
\cline{2-9}
& $\tilde{M}=0$ & ---
& $\tilde{M}=0$ & rIV 
& $\tilde{M}=0$ & rII 
& $\tilde{M}=0$ & rII
\\
\cline{2-9}
& $\tilde{M}>0$ & sVII
& $\tilde{M}>0$ & sVIII 
& $\tilde{M}>0$ & bIII 
\\
\noalign{\hrule height 1.0pt}
-1 &  $\tilde{M}<\tilde{M}_{B}$ & sIII
& $\tilde{M}<\tilde{M}_{B}$ & sII 
& $\tilde{M}<\tilde{M}_{ex}^{(-)}$ & sIV 
& $\tilde{M}<\tilde{M}_{B}$ & sIV
\\
\cline{2-9}
& $\tilde{M}=\tilde{M}_{B}$ & sVII
& $\tilde{M}=\tilde{M}_{B}$ & sVIII
& $\tilde{M}=\tilde{M}_{ex}^{(-)}$ & dbI 
& $\tilde{M}=\tilde{M}_{B}$ & sIV
\\
\cline{2-9}
& $\tilde{M}>\tilde{M}_{B}$ $(\tilde{M}\ne 0)$ & sVII
& $\tilde{M}>\tilde{M}_{B}$ $(\tilde{M}\ne 0)$ & sVIII 
& $\tilde{M}_{ex}^{(-)}<\tilde{M}<\tilde{M}_{B}$ & bIV   
& $\tilde{M}_{B}<\tilde{M}<\tilde{\alpha}$ $(\tilde{M}\ne 0)$ & bIII
\\
\cline{2-9}
& $\tilde{M}=0$ & rIII
& $\tilde{M}=0$ & rVI 
& $\tilde{M}=\tilde{M}_{B}$ & bIII 
& $\tilde{M}=0$ & bV
\\
\cline{2-9}
&  & 
&  & 
&  $\tilde{M}>\tilde{M}_{B}$ $(\tilde{M}\ne 0)$ & bIII
& $\tilde{M}=\tilde{\alpha}$ & sVI
\\
\cline{6-9}
&  & 
&  & 
& $\tilde{M}=0$  & bV
& $\tilde{M}>\tilde{\alpha}$ & sIV
\\
\hline
\hline
\end{tabular}
\end{table}



\begin{table}
\renewcommand{\arraystretch}{1.2}
\caption{ 
The classification of the spacetime structures of the static solutions in a Einstein-Gauss-Bonnet-$\Lambda$
system where $4\tilde{\alpha}/\ell^2=1$.  The
numbers in the column ``type" imply the types of the conformal diagrams in Figs.~\ref{penrose_regular}-\ref{penrose_BH_degenerate}. }
\label{GB-penrose-spe-neutral}
\vspace{4mm}
\begin{tabular}{c|c|c|c|c|c|c|c|c}
\hline
\hline
&\multicolumn{4}{c|}{$n\geq 6$} 
&\multicolumn{4}{c}{$n=5$}
\\
\cline{2-9}
~$k$~
&\multicolumn{2}{c|}{$-$ branch} 
&\multicolumn{2}{c|}{$+$ branch} 
&\multicolumn{2}{c|}{$-$ branch}
&\multicolumn{2}{c}{$+$ branch}
\\
\cline{2-9}
&  ~~~~~~~~~~$\tilde{M}$~~~~~~~~~~ & ~type~ 
&  ~~~~~~~~~~$\tilde{M}$~~~~~~~~~~ & ~type~ 
&  ~~~~~~~~~~$\tilde{M}$~~~~~~~~~~ & ~type~ 
&  ~~~~~~~~~~$\tilde{M}$~~~~~~~~~~ & ~type~ 
\\
\hline \hline 
1
& $\tilde{M}=0$ & rVII
& $\tilde{M}=0$ & rVII  
& $\tilde{M}=0$ & rVII
& $\tilde{M}=0$ & rVII  
\\
\cline{2-9}
& $\tilde{M}>0$ & bIII
& $\tilde{M}>0$ & sIV
& $0<\tilde{M}<\tilde{\alpha}$ & sIV
& $\tilde{M}>0$ & sIV
\\
\cline{2-9}
&  & 
&  & 
& $\tilde{M}=\tilde{\alpha}$ & sVI 
\\
\cline{6-7}
&  & 
&  & 
& $\tilde{M}>\tilde{\alpha}$ & bIII 
\\
\noalign{\hrule height 1.0pt}
0 
& $\tilde{M}=0$ & rII
& $\tilde{M}=0$ & rII 
& $\tilde{M}=0$ & rII
& $\tilde{M}=0$ & rII
\\
\cline{2-9}
& $\tilde{M}>0$ & bIII
& $\tilde{M}>0$ & sIV
& $0<\tilde{M}<\tilde{\alpha}$ & bIII
& $\tilde{M}>0$ & sIV
\\
\noalign{\hrule height 1.0pt}
-1 
& $\tilde{M}=0$ & bV
& $\tilde{M}=0$ & bV
& $\tilde{M}=0$ & bV
& $\tilde{M}=0$ & bV 
\\
\cline{2-9}
& $\tilde{M}>0$ & bIII
& $0<\tilde{M}<\tilde{M}_{ex}^{(+)}$ & bIV 
& $\tilde{M}>0$ & bIII 
& $0<\tilde{M}<\tilde{\alpha}$ & bIII
\\
\cline{2-9}
&   &  
& $\tilde{M}=\tilde{M}_{ex}^{(+)}$ & dbI 
&   &  
& $\tilde{M}=\tilde{\alpha}$ & sVI
\\
\cline{2-9}
&   &  
& $\tilde{M}>\tilde{M}_{ex}^{(+)}$ & sIV 
&   &  
& $\tilde{M}>\tilde{\alpha}$ & sIV
\\
\hline
\hline
\end{tabular}
\end{table}

\end{widetext}



\begin{references}
\bibitem{regular} 
   When the $(n-2)$-dimensional Einstein space is simply
connected, the center is regular. When it is not, as with the topological black 
holes, the center becomes singular. This singularity is simply due to the
topological structure of the Einstein space. Hence we use the word
``regular" at the center, even in this case.
\bibitem{note_1} 
   As we mentioned, in the higher-dimensional cases, the $(n-2)$-dimensional
Einstein space is not necessarily a sphere nor is its quotient space even
for positive curvature. Hence the solution represents not only the (at least locally) Minkowski spacetime but also
many others. However, here and in the following sections we refer to all of these together as 
``Minkowski spacetimes".
\end{references}
\end{document}